\documentclass[letterpaper,twoside,twocolumn,english,superscriptaddress,unsortedaddress,secnumarabic,nobibnotes,final,prl]{revtex4}
\usepackage[latin9]{inputenc}
\usepackage{babel}
\usepackage{units}
\usepackage{amsbsy}
\usepackage{amstext}
\usepackage{graphicx}
\usepackage[unicode=true,
 bookmarks=true,bookmarksnumbered=false,bookmarksopen=false,
 breaklinks=false,pdfborder={0 0 1},backref=false,colorlinks=false]
 {hyperref}
\hypersetup{pdftitle={A Machine Learning Potential For Large-Scale Simulations Of Carbon At Extreme Conditions},
 pdfauthor={Jonathan Willman},
 pdfsubject={Machine Learning Potential}}

\makeatletter

\pdfpageheight\paperheight
\pdfpagewidth\paperwidth

\@ifundefined{textcolor}{}
{%
 \definecolor{BLACK}{gray}{0}
 \definecolor{WHITE}{gray}{1}
 \definecolor{RED}{rgb}{1,0,0}
 \definecolor{GREEN}{rgb}{0,1,0}
 \definecolor{BLUE}{rgb}{0,0,1}
 \definecolor{CYAN}{cmyk}{1,0,0,0}
 \definecolor{MAGENTA}{cmyk}{0,1,0,0}
 \definecolor{YELLOW}{cmyk}{0,0,1,0}
}

\makeatother

\begin{document}

\title{Machine Learning Interatomic Potential for Simulations of Carbon
at Extreme Conditions}

\author{Jonathan T. Willman}

\affiliation{University of South Florida, Tampa, Florida 33620, USA}

\author{Kien Nguyen-Cong}

\affiliation{University of South Florida, Tampa, Florida 33620, USA}

\author{Ashley S. Williams}

\affiliation{University of South Florida, Tampa, Florida 33620, USA}

\author{Anatoly B. Belonoshko}

\affiliation{Royal Institute of Technology, 106691 Stockholm, Sweden}

\author{Stan G. Moore}

\affiliation{Sandia National Laboratories, Albuquerque, New Mexico 87185, USA}

\author{Aidan P. Thompson}

\affiliation{Sandia National Laboratories, Albuquerque, New Mexico 87185, USA}

\author{Mitchell A. Wood}

\affiliation{Sandia National Laboratories, Albuquerque, New Mexico 87185, USA}

\author{Ivan I. Oleynik}
\email{oleynik@usf.edu}

\affiliation{University of South Florida, Tampa, Florida 33620, USA}
\begin{abstract}
A Spectral Neighbor Analysis (SNAP) machine learning interatomic potential
(MLIP) has been developed for simulations of carbon at extreme pressures
(up to 5 TPa) and temperatures (up to 20,000 K). This was achieved
using a large database of experimentally relevant quantum molecular
dynamics (QMD) data, training the SNAP potential using a robust machine
learning methodology, and performing extensive validation against
QMD and experimental data. The resultant carbon MLIP demonstrates
unprecedented accuracy and transferability in predicting the carbon
phase diagram, melting curves of crystalline phases, and the shock
Hugoniot, all within $\unit[3]{\%}$ of QMD. By achieving quantum
accuracy and efficient implementation on leadership class high performance
computing systems, SNAP advances frontiers of classical MD simulations
by enabling atomic-scale insights at experimental time and length
scales.
\end{abstract}
\maketitle
Carbon at extreme pressures and temperatures is a topic of great scientific
interest for several disciplines including planetary science \citep{Ross1981,Ancilotto1997,Benedetti1999,Kraus2017,Madhusudhan2012}
and inertial confinement fusion (ICF) research \citep{Ross2015,Millot2018,Hopkins2019,Kritcher2021}.
Methane ice at megabar pressures and temperatures of thousands of
kelvins is predicted to convert to solid or liquid carbon in the cores
of giant planets \citep{Ancilotto1997,Benedetti1999,Kraus2017}. A
successful suppression of hydrodynamic instabilities seeded by solid
and liquid carbon phases appearing upon strong compression of the
outer diamond ablation shell of an ICF capsule \citep{Casey2021}
was the key for achieving a record-breaking, near-threshold fusion
energy ignition at the National Ignition Facility \citep{Tollefson2021}. 

The exploration of carbon's behavior at extreme conditions is challenging
for both theory and experiment. Shock and ramp compression experiments
using powerful lasers \citep{Duffy2019}, pulsed power \citep{Sinars2020}
and bright X-ray sources \citep{McMahon2020,Seddon2017} uncovered
complicated mechanisms of inelastic deformations \citep{Pavlovskii1971,Kondo1983,McWilliams2010,Lang2018,Katagiri2020,Winey2020},
anomalous strength of diamond \citep{McWilliams2010,Lang2018,Katagiri2020,Winey2020,MacDonald2020},
unusual melting \citep{Brygoo2007,Knudson2008,Eggert2009,Gregor2017,Millot2018,Millot2020}
and liquid carbon properties \citep{Knudson2008,Eggert2009}, as well
as extreme metastability of diamond well beyond the pressure-temperature
range of its thermodynamic stability \citep{Lazicki2021}. Molecular
dynamics (MD) simulations can provide a fundamental understanding
of these phenomena, but to be of experimental relevance, it must accurately
describe interatomic interactions in a system consisting of a large
number of atoms. 

Previous simulations of carbon at extreme conditions were predominantly
performed using quantum molecular dynamics (QMD) based on density
functional theory (DFT) \citep{Scandolo1996,Grumbach1996,Wang2005,Correa2006,Correa2008,Knudson2008,Sun2009,Martinez-Canales2012,Benedict2014}.
Due to high computational cost, QMD simulations are limited to several
hundred atoms for up to tens of picoseconds, which is insufficient
for uncovering non-equilibrium processes at experimental time (ns)
and length ($\mathrm{\mathrm{\mu}}m$) scales. In principle, these
scales can be accessed by classical MD simulations on massively parallel
computers \citep{Thompson2022}. However, empirical interatomic potentials
developed for carbon at near ambient conditions \citep{Brenner2002,Stuart2000,Los2003,Pastewka2008,Srinivasan2015}
singularly fail upon extension to high pressures and temperatures
\citep{Oleynik2008,Perriot2013}, thus compromising predictive power
of atomistic simulations. 

The advent of machine learning interatomic potentials (MLIPs) \citep{Behler2007,Bartok2010}
opens up exceptional opportunities for achieving a classical description
of chemical bonding with quantum accuracy \citep{Friederich2021}.
Although numerous MLIPs have been recently introduced and successfully
applied to modeling properties of materials at ambient conditions
\citep{Behler2007,Bartok2010,Thompson2015,Wood2018,Shapeev2015,Drautz2019,Zhang2018},
including carbon \citep{Khaliullin2010,Rowe2020}, their exceptional
power in describing diverse and complex atomic environments at megabar
pressures and tens of thousand of kelvins has yet to be demonstrated. 

This letter reports a significant advance in development of a quantum-accurate
Spectral Neighbor Analysis Potential (SNAP) for simulations of carbon
at extreme pressure-temperature (P-T) conditions. This includes construction
of an experimentally relevant training database, implementation of
a robust SNAP machine learning training methodology, and extensive
validation against QMD and experimental data. The end result is the
first MLIP that delivers unprecedented accuracy in simulating carbon
over a remarkably wide range of pressures (from 0 to 50 Mbar) and
temperatures (up to 20,000 K).

\begin{figure*}[!t]
\includegraphics[width=6.8in]{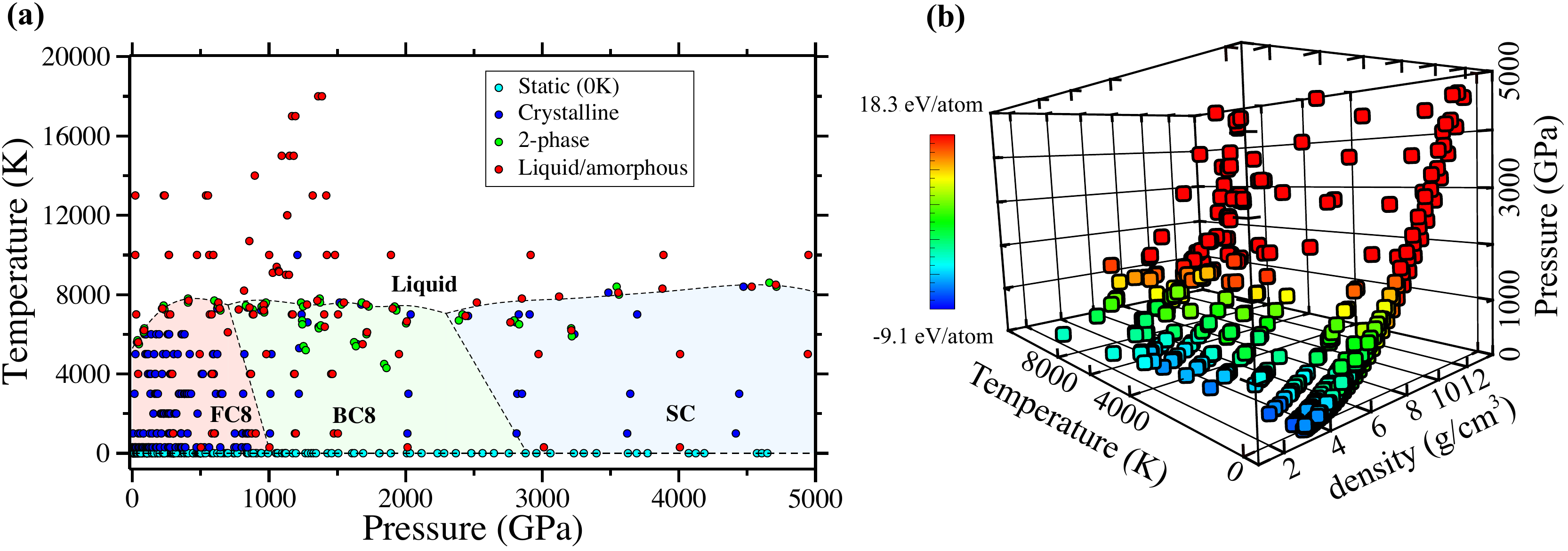}\caption{SNAP training database: a) pressure-temperature map of QMD and static
DFT simulations included in the database, each represented by a P-T
point on carbon phase diagram sampling FC8 (diamond), BC8, SC solid
and liquid phases (total number of structures - $636$); b) pressure-temperature-density-energy/atom
distribution. }
\end{figure*}

In general, MLIPs fingerprint a unique local atomic environment around
each atom by a set of descriptors. SNAP's descriptors are bispectrum
components $\{\mathbf{B^{\mathit{i}}}\}$ of the local neighbor density
projected onto a basis of hyper-spherical harmonics in four dimensions
\citep{Thompson2015,Wood2018}. Other successful MLIPs -- Neural
Network Potentials (NNP) \citep{Behler2007,Zhang2018}, Gaussian Approximation
Potential (GAP) \citep{Bartok2010}, the Moment Tensor Potential (MTP)
\citep{Shapeev2015} -- employ mathematically different, but physically
similar descriptors. All of them, including SNAP can be mapped into
a general Atomic Cluster Expansion (ACE) descriptor framework \citep{Drautz2019}. 

Herein, the total potential energy of the system of $N$ atoms is
written as a sum of atomic energies $E^{i}$, which are quadratic
functions of the bispectrum coefficients $\mathbf{B^{\mathit{i}}}$
\citep{Wood2018}

\begin{equation}
E_{tot}(\{\mathbf{r}^{N}\})=\sum_{i}E^{i};\qquad E^{i}=\mathbf{\boldsymbol{\boldsymbol{\beta}}\cdotp B^{\mathit{i}}+}\frac{1}{2}\mathbf{B^{\mathit{i}}\cdotp\boldsymbol{\alpha}\cdotp B^{\mathit{i}}}.\label{eq:snap}
\end{equation}
Machine-learning techniques are used to determine the symmetric matrix
$\boldsymbol{\boldsymbol{\alpha}}$ and the vector $\boldsymbol{\boldsymbol{\beta}}$,
the unknown parameters of SNAP, to reproduce potential energy, atomic
forces and stress tensor for each structure in the DFT training database. 

SNAP displays a good balance between computational cost and accuracy
\citep{Ong2020}. Both are controlled by SNAP hyperparameters --
the cutoff radius $r_{cut}$ and the integer angular momentum $J$.
The former specifies the number of neighbor atoms participating in
the fingerprinting of atomic environment around each atom $i$ and
the latter refers to the dimensionality of the descriptor space, i.e.
the number of bispectrum descriptors$\{\mathbf{B^{\mathit{i}}}\}$:
$(J+1)(J+3/2)(J+2)/2$ for each atom $\mathit{i}$.

The SNAP development includes: (i) construction of a robust training
database of first-principles QMD data; (ii) machine-learning training;
and (iii) extensive validation against QMD and experiment. These three
tightly connected steps constitute a single development cycle. Several
of such cycles are executed to improve upon deficiencies observed
in previous iterations, see the development workflow in supplemental
Fig. S2 \citep{S_material}. For example, during the validation, two-phase
SNAP MD simulations produce melting lines of several high-pressure
carbon phases in a substantial disagreement with QMD. The problem
has been traced back to SNAP inaccuracies in calculation of enthalpies
of solid and liquid phases along the melting line, which, according
to Clausius--Clapeyron relation, define its slope $dT/dP$. Therefore,
separate liquid and solid phases were added to the QMD training database
to complement original combined two-phase solid-liquid structures.
This update resulted in substantial improvements in SNAP accuracy
upon execution of a new training cycle.

\begin{figure*}
\includegraphics[width=6.8in]{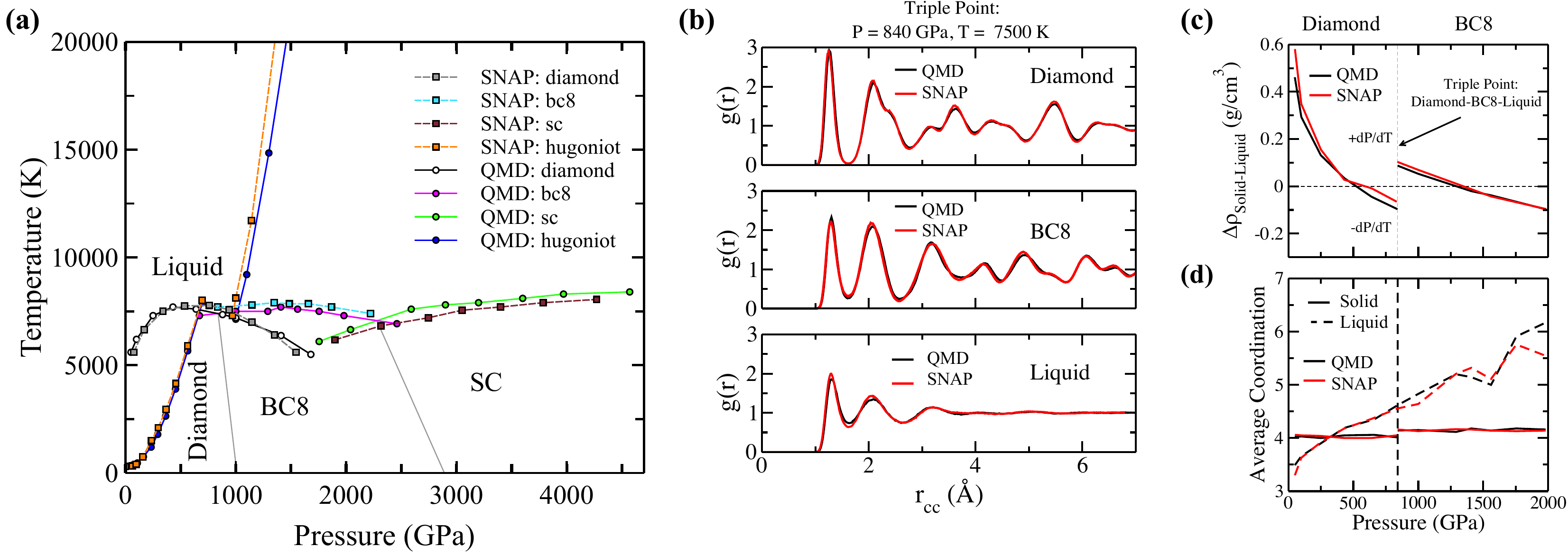}

\caption{Validation of SNAP against QMD: (a) carbon phase diagram, including
melting lines of diamond, BC8 and SC crystalline phases of carbon
at pressures up to $5,000$ GPa and hydrostatic shock Hugoniot; (b)
Radial distribution functions $g(r)$ for diamond, BC8 and liquid
phase at diamond-bc8-liquid triple point ($840$ GPa, $7510$ K).
(MD trajectories are averaged over $20$ ps time interval); c) density
difference $\Delta\rho=\rho_{s}-\rho_{l}$ between solid (s) and liquid
(l) phases (top panel) and average coordination number of carbon atoms
(bottom panel) as a function of pressure along the melting lines of
diamond and BC8. $\Delta\rho=0$ corresponds to melting line maxima:
diamond -- at $\unit[500]{GPa}$, BC8 -- at $\unit[1,300]{GPa}$.}
\end{figure*}

One of the highlights of our MLIP development is the dedicated construction
of a SNAP training database of experimentally relevant QMD of large
systems (by QMD standards) -- up to 1,024 atoms. These are individual
frames (1-3 frames per simulation) from QMD production simulations
of physical properties of carbon (melting lines, hydrostatic and uniaxial
isotherms, shock Hugoniot) performed within a wide range of pressures
(from $0$ to $5$ TPa), temperatures (from $0$ to $40,000$ K) and
densities (from $2.9$ to $13.6$ $\unit{g/cm^{3}}$). For example,
MD frames for each $(P,T)$ point along the melting lines of several
crystalline carbon phases were taken from production two-phase QMD
simulation \citep{Belonoshko1994}. These complex cells contain both
liquid and solid parts separated by a realistic solid-liquid interface,
which add an additional complexity to the SNAP database. A series
of complimentary QMD simulations of liquid and solid phases were also
included in the database. The QMD data is supplemented by static DFT
calculations of binding energy curves, point and extended defects,
and metastable carbon structures obtained from dedicated crystal structure
searching \citep{Williams2022}. Fig.~1 displays the range of pressures,
temperatures and energies covered by QMD simulations. The SNAP database,
consisting of $636$ structures, samples 124,907 unique atomic environments.
Each structure of $N$ atoms contributes $1$ total energy, $6$ stress
components and $3N$ atomic forces resulting in a total training complexity
(the number of regression equations to fit) of $388,077$. Additional
information on database composition and its generation is provided
in the Supplemental Material \citep{S_material}. 

Once the database is constructed, the SNAP model for the total potential
energy, stress tensor components and atomic forces is trained using
machine learning techniques to determine SNAP parameters $\boldsymbol{\alpha}$
and $\boldsymbol{\beta}$ through minimization of an objective function
- a sum of the normalized squared differences between DFT and SNAP
energies, stresses and atomic forces (Fig. S2) \citep{S_material}.
The training is performed in a series of iterations. For a given set
of weights, the SNAP parameters $\boldsymbol{\boldsymbol{\alpha}}$
and $\boldsymbol{\beta}$ are determined through weighted linear regression
as implemented in FitSNAP package \citep{fitsnap}. The weights are
then optimized to minimize the objective function using a genetic
algorithm (GA) within DAKOTA software package \citep{Adams}. At every
step of GA minimization FitSNAP is called with the current set of
weights to determine new $\boldsymbol{\boldsymbol{\alpha}}$ and $\mathbf{\boldsymbol{\beta}}$,
which are then fed back to update the objective function being minimized
(Fig. S2) \citep{S_material}. The iterations stop when the GA minimization
converges to a final set of weights. In addition to group weights,
the SNAP hyperparameters, $r_{cut}$ and $J$ are optimized in an
offset fashion to find a right balance between SNAP accuracy and computational
efficiency (Fig. S2 \citep{S_material}). The final values for the
SNAP hyperparameters are $r_{cut}=\unit[2.7]{\text{Å}}$ and $J=4$.
The resultant quality of SNAP training is discussed in Supplemental
Material \citep{S_material}. 

The critical part of SNAP potential development is a thorough validation
of SNAP MD results against QMD and experimental data. Only one to
three QMD frames per (P,T) state were included in the SNAP training
database. Therefore, simulating these states with SNAP using much
larger simulations cells and performing thermodynamic averaging of
atomic trajectories containing tens of thousands of frames to obtain
stresses, densities, internal energies, as well as radial distribution
functions is considered as a rigorous validation test against QMD.
Further, these validation simulations are extended to sample a variety
of pressures and temperatures not included in the database and compared
to both QMD and experiment to demonstrate SNAP transferability.

The important validation test is concerned with carbon phase diagram,
including melting lines of several high-pressure phases at pressures
up to $5,000$ GPa, see Fig. (2a). A series of isobaric-isothermal
NPT two-phase MD simulations were run at a given pressure but varying
temperature to determine the P-T value of the phase coexistence \citep{Belonoshko1994}.
SNAP melting lines are in excellent agreement with those from QMD,
the average temperature error being $\sim\unit[200]{K}$ or $3\%$
in a pressure interval from $0$ to $5,000$ GPa. Fig. 2(b) displays
SNAP and QMD radial distribution functions $g(r)$ at diamond-BC8-liquid
triple point: they are almost indistinguishable from each other.

A remarkable property of carbon at extreme conditions is the negative
slope ($dT/dP$) of diamond and BC8 melting lines at high pressures
\citep{Grumbach1996,Wang2005,Correa2006,Correa2008,Benedict2014,Eggert2009}.
This is because liquid carbon becomes denser than the corresponding
solid phase upon increase of pressure above $\sim500$ GPa for diamond
and $\sim1,300$ GPa for BC8, see Fig. 2(c). This can be traced back
to a significant increase of carbon packing in the liquid as carbon
atom coordination changes from less than 4 to higher values, see Fig.
2(d). SNAP accurately predicts this subtle change in density upon
increase of pressure as well as corresponding pressure-dependent evolution
of the average coordination of carbon atoms ( Fig. 2(d)), which is
in an excellent agreement with QMD. The latter is a result of ``snap-on''
agreement between QMD and SNAP radial distribution functions for both
diamond and BC8 over a large range of pressures (Fig. S4 \citep{S_material}).

\begin{figure}
\includegraphics[width=0.9\columnwidth]{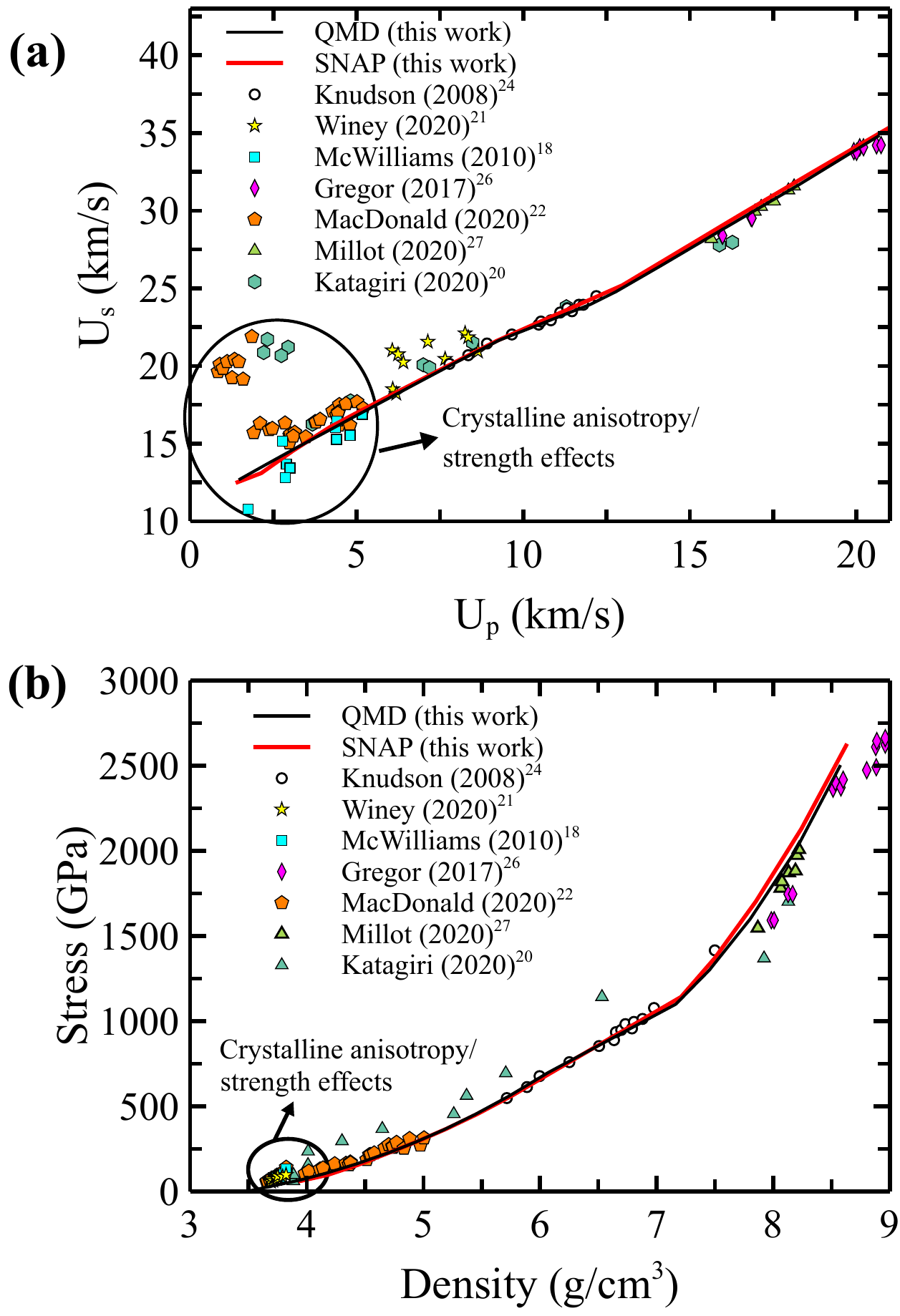}

\caption{Carbon shock Hugoniot calculated by QMD and SNAP and compared with
experimental data. (a) $U_{s}-U_{p}$ Hugoniot (b) pressure-density
Hugoniots. Points correspond to experimental data.}
\end{figure}

Another validation test of great experimental importance is the prediction
of the carbon shock Hugoniot, which passes through both solid and
liquid parts of the phase diagram (Fig. 2(a)). The Hugoniot points
are calculated in a hydrodynamic approximation by ignoring crystalline
anisotropy in a series of MD simulations at a given pressure $P$
but varying temperature $T$ to satisfy the Hugoniot condition of
conservation of mass, momentum, and energy: $\frac{1}{2}(P+P_{0})(V-V_{0})=(E-E_{0})$,
where $P$, $V,$ and $E$ are the pressure, volume, and internal
energy at a given point on the Hugoniot, and $P_{0}$, $V_{0}$, and
$E_{0}$ are those at ambient conditions of $0$ GPa and $300$ K.
The Hugoniot in $(P,T)$ space is shown in Fig. 2(a), in pressure-density
$(P-\rho)$ space -- in Fig. 3(b) and particle velocity $U_{p}$
-- shock velocity $U_{s}$ space -- in Fig. 3(a). The SNAP Hugoniots
(red lines in Figs 2(a) and 3) are in a very good agreement with those
from QMD (black lines in Figs 2(a) and 3). Visible differences in
temperature of the \emph{P--T} Hugoniot at very high temperatures
(Fig. 2(a)) are due to electronic entropy effects, which are not captured
by any classical interatomic potential. 

Both SNAP and QMD Hugoniots are also in a very good agreement with
multiple experiments in a pressure range from $300$ to $\unit[1,500]{GPa}$
\citep{Knudson2008,McWilliams2010,Gregor2017,MacDonald2020,Millot2020,Winey2020,Katagiri2020}.
Some difference between SNAP/QMD and experiment at higher pressures
is due to experimental uncertainty in density, which is not measured
directly, but rather determined using an impedance matching method
\citep{Gregor2017,Millot2020}. Differences at low pressures between
experiment \citep{McWilliams2010,MacDonald2020,Winey2020} and SNAP/QMD
(Fig.~3(a)) are due to crystalline anisotropy and strength effects,
which are not well described by a hydrodynamic approximation \citep{Winey2020}.
To make a proper comparison with experiment in this split elastic-inelastic
shock wave regime, explicit large-scale SNAP MD simulations of piston-driven
shock waves are required, which will be the focus of future work.

\begin{figure*}[!t]
\includegraphics[width=6.8in]{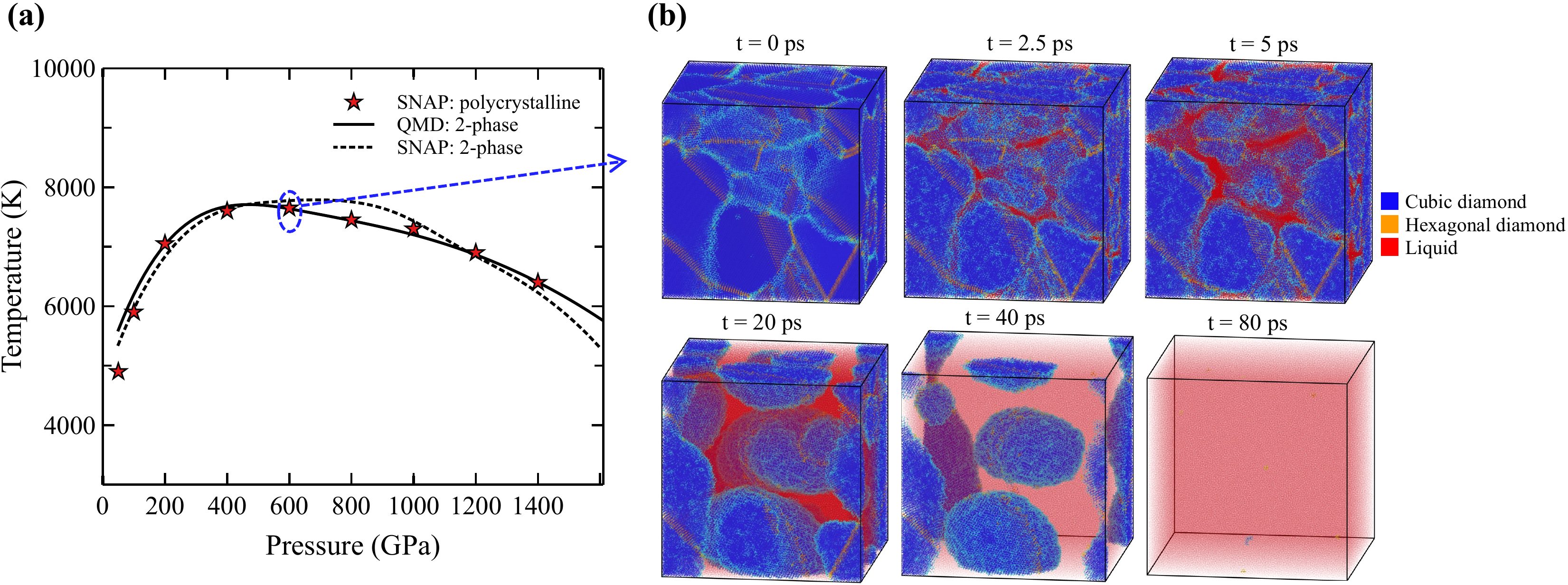}

\caption{(a) Comparison of diamond melting curves calculated by three methods:
QMD 2-phase (solid line), SNAP 2-phase (dashed line), and direct melting
of 1 million atom polycrystalline sample using SNAP (stars); (b) Time
progression of polycrystalline sample melting at $600$ GPa and $7750$
K. The sample is initially composed of crystalline diamond regions
(blue) separated by grain boundaries (light blue). Liquid regions
(red) emerge at the grain boundaries, grow in size, and eventually
consume the isolated diamond crystallites. }
\end{figure*}

To demonstrate SNAP's ability to attack problems that are impossible
to solve with QMD, we simulated melting of polycrystalline diamond
at pressures between 50 and 1,200 GPa using a 1 million atom sample
(Fig. 4). For each pressure point, a series of NPT simulations is
performed to determine the temperature at the onset of melting. In
addition to validating SNAP, this simulation also validates the two-phase
melting curve calculation. The presence of grain boundaries suppresses
superheating characteristic of single crystals: the melting starts
at the most weakly bonded defective regions of the sample, followed
by the growth of liquid fraction at the expense of crystalline grains,
which gradually transform to shrinking round crystallites embedded
in the liquid carbon (Fig. 4).

This work represents a major step towards solving extremely challenging
but fundamentally important problem of predictive atomic-scale simulations
of carbon at extreme pressure-temperature conditions at experimental
time and length scales. The quantum-accurate SNAP is the first MLIP
that describes the properties of carbon at extreme pressures up to
$5$ TPa and temperatures up to $20,000$ K, including the phase diagram,
melting curves of diamond, BC8 and simple cubic phases of carbon and
shock Hugoniots with unprecedented accuracy within $\unit[3]{\%}$
of QMD results. SNAP's linear scaling with number of atoms, and its
efficient implementation within the LAMMPS MD simulation package \citep{Thompson2022}
allows billion atom simulations on leadership class high performance
computing systems \citep{Nguyen-Cong2021}. By advancing frontiers
of classical MD simulations, SNAP enables new insights by uncovering
fundamental atomic-scale mechanisms of materials response which are
difficult or even impossible to observe in experiment.

The work at USF is supported by DOE/NNSA (grant DE-NA- 0003910). The
work at KTH is supported by Swedish Scientific Council (VR) (grant
2017-03744). Sandia National Laboratories is a multi-mission laboratory
managed and operated by National Technology and Engineering Solutions
of Sandia, LLC, a wholly owned subsidiary of Honeywell International,
Inc., for the U.S. Department of Energy\textquoteright s (DOE) National
Nuclear Security Administration under Contract No. DE-NA0003525. The
computations were performed using leadership-class HPC systems: OLCF
Summit at Oak Ridge National Laboratory (ALCC and INCITE awards MAT198)
and TACC Frontera at University of Texas at Austin (LRAC award DMR21006).

\bibliographystyle{apsrev4-1}
\bibliography{SNAP-C-ArXiv-05-02-2022}

\begin{thebibliography}{61}%
\makeatletter
\providecommand \@ifxundefined [1]{%
 \@ifx{#1\undefined}
}%
\providecommand \@ifnum [1]{%
 \ifnum #1\expandafter \@firstoftwo
 \else \expandafter \@secondoftwo
 \fi
}%
\providecommand \@ifx [1]{%
 \ifx #1\expandafter \@firstoftwo
 \else \expandafter \@secondoftwo
 \fi
}%
\providecommand \natexlab [1]{#1}%
\providecommand \enquote  [1]{``#1''}%
\providecommand \bibnamefont  [1]{#1}%
\providecommand \bibfnamefont [1]{#1}%
\providecommand \citenamefont [1]{#1}%
\providecommand \href@noop [0]{\@secondoftwo}%
\providecommand \href [0]{\begingroup \@sanitize@url \@href}%
\providecommand \@href[1]{\@@startlink{#1}\@@href}%
\providecommand \@@href[1]{\endgroup#1\@@endlink}%
\providecommand \@sanitize@url [0]{\catcode `\\12\catcode `\$12\catcode
  `\&12\catcode `\#12\catcode `\^12\catcode `\_12\catcode `\%12\relax}%
\providecommand \@@startlink[1]{}%
\providecommand \@@endlink[0]{}%
\providecommand \url  [0]{\begingroup\@sanitize@url \@url }%
\providecommand \@url [1]{\endgroup\@href {#1}{\urlprefix }}%
\providecommand \urlprefix  [0]{URL }%
\providecommand \Eprint [0]{\href }%
\providecommand \doibase [0]{http://dx.doi.org/}%
\providecommand \selectlanguage [0]{\@gobble}%
\providecommand \bibinfo  [0]{\@secondoftwo}%
\providecommand \bibfield  [0]{\@secondoftwo}%
\providecommand \translation [1]{[#1]}%
\providecommand \BibitemOpen [0]{}%
\providecommand \bibitemStop [0]{}%
\providecommand \bibitemNoStop [0]{.\EOS\space}%
\providecommand \EOS [0]{\spacefactor3000\relax}%
\providecommand \BibitemShut  [1]{\csname bibitem#1\endcsname}%
\let\auto@bib@innerbib\@empty
\bibitem [{\citenamefont {Ross}(1981)}]{Ross1981}%
  \BibitemOpen
  \bibfield  {author} {\bibinfo {author} {\bibfnamefont {M.}~\bibnamefont
  {Ross}},\ }\href@noop {} {\bibfield  {journal} {\bibinfo  {journal} {Nature}\
  }\textbf {\bibinfo {volume} {292}},\ \bibinfo {pages} {435} (\bibinfo {year}
  {1981})}\BibitemShut {NoStop}%
\bibitem [{\citenamefont {Ancilotto}\ \emph {et~al.}(1997)\citenamefont
  {Ancilotto}, \citenamefont {Chiarotti}, \citenamefont {Scandolo},\ and\
  \citenamefont {Tosatti}}]{Ancilotto1997}%
  \BibitemOpen
  \bibfield  {author} {\bibinfo {author} {\bibfnamefont {F.}~\bibnamefont
  {Ancilotto}}, \bibinfo {author} {\bibfnamefont {G.~L.}\ \bibnamefont
  {Chiarotti}}, \bibinfo {author} {\bibfnamefont {S.}~\bibnamefont {Scandolo}},
  \ and\ \bibinfo {author} {\bibfnamefont {E.}~\bibnamefont {Tosatti}},\ }\href
  {\doibase 10.1126/science.275.5304.1288} {\bibfield  {journal} {\bibinfo
  {journal} {Science}\ }\textbf {\bibinfo {volume} {275}},\ \bibinfo {pages}
  {1288} (\bibinfo {year} {1997})}\BibitemShut {NoStop}%
\bibitem [{\citenamefont {Benedetti}\ \emph {et~al.}(1999)\citenamefont
  {Benedetti}, \citenamefont {Nguyen}, \citenamefont {Caldwell}, \citenamefont
  {Liu}, \citenamefont {Kruger},\ and\ \citenamefont
  {Jeanloz}}]{Benedetti1999}%
  \BibitemOpen
  \bibfield  {author} {\bibinfo {author} {\bibfnamefont {L.~R.}\ \bibnamefont
  {Benedetti}}, \bibinfo {author} {\bibfnamefont {J.~H.}\ \bibnamefont
  {Nguyen}}, \bibinfo {author} {\bibfnamefont {W.~A.}\ \bibnamefont
  {Caldwell}}, \bibinfo {author} {\bibfnamefont {H.}~\bibnamefont {Liu}},
  \bibinfo {author} {\bibfnamefont {M.}~\bibnamefont {Kruger}}, \ and\ \bibinfo
  {author} {\bibfnamefont {R.}~\bibnamefont {Jeanloz}},\ }\href@noop {}
  {\bibfield  {journal} {\bibinfo  {journal} {Science}\ }\textbf {\bibinfo
  {volume} {286}} (\bibinfo {year} {1999})}\BibitemShut {NoStop}%
\bibitem [{\citenamefont {Kraus}\ \emph {et~al.}(2017)\citenamefont {Kraus},
  \citenamefont {Vorberger}, \citenamefont {Pak}, \citenamefont {Hartley} \emph
  {et~al.}}]{Kraus2017}%
  \BibitemOpen
  \bibfield  {author} {\bibinfo {author} {\bibfnamefont {D.}~\bibnamefont
  {Kraus}}, \bibinfo {author} {\bibfnamefont {J.}~\bibnamefont {Vorberger}},
  \bibinfo {author} {\bibfnamefont {A.}~\bibnamefont {Pak}}, \bibinfo {author}
  {\bibfnamefont {N.~J.}\ \bibnamefont {Hartley}},  \emph {et~al.},\
  }\href@noop {} {\bibfield  {journal} {\bibinfo  {journal} {Nat. Astron.}\
  }\textbf {\bibinfo {volume} {1}},\ \bibinfo {pages} {606} (\bibinfo {year}
  {2017})}\BibitemShut {NoStop}%
\bibitem [{\citenamefont {Madhusudhan}\ \emph {et~al.}(2012)\citenamefont
  {Madhusudhan}, \citenamefont {Lee},\ and\ \citenamefont
  {Mousis}}]{Madhusudhan2012}%
  \BibitemOpen
  \bibfield  {author} {\bibinfo {author} {\bibfnamefont {N.}~\bibnamefont
  {Madhusudhan}}, \bibinfo {author} {\bibfnamefont {K.~K.~M.}\ \bibnamefont
  {Lee}}, \ and\ \bibinfo {author} {\bibfnamefont {O.}~\bibnamefont {Mousis}},\
  }\href@noop {} {\bibfield  {journal} {\bibinfo  {journal} {Astrophys. J.}\
  }\textbf {\bibinfo {volume} {759}},\ \bibinfo {pages} {L40} (\bibinfo {year}
  {2012})}\BibitemShut {NoStop}%
\bibitem [{\citenamefont {Ross}\ \emph {et~al.}(2015)\citenamefont {Ross},
  \citenamefont {Ho}, \citenamefont {Milovich}, \citenamefont {D{\"{o}}ppner},
  \citenamefont {McNaney} \emph {et~al.}}]{Ross2015}%
  \BibitemOpen
  \bibfield  {author} {\bibinfo {author} {\bibfnamefont {J.~S.}\ \bibnamefont
  {Ross}}, \bibinfo {author} {\bibfnamefont {D.}~\bibnamefont {Ho}}, \bibinfo
  {author} {\bibfnamefont {J.}~\bibnamefont {Milovich}}, \bibinfo {author}
  {\bibfnamefont {T.}~\bibnamefont {D{\"{o}}ppner}}, \bibinfo {author}
  {\bibfnamefont {J.}~\bibnamefont {McNaney}},  \emph {et~al.},\ }\href@noop {}
  {\bibfield  {journal} {\bibinfo  {journal} {Phys. Rev. E}\ }\textbf {\bibinfo
  {volume} {91}},\ \bibinfo {pages} {1} (\bibinfo {year} {2015})}\BibitemShut
  {NoStop}%
\bibitem [{\citenamefont {Millot}\ \emph {et~al.}(2018)\citenamefont {Millot},
  \citenamefont {Celliers}, \citenamefont {Sterne}, \citenamefont {Benedict},
  \citenamefont {Correa}, \citenamefont {Hamel}, \citenamefont {Ali} \emph
  {et~al.}}]{Millot2018}%
  \BibitemOpen
  \bibfield  {author} {\bibinfo {author} {\bibfnamefont {M.}~\bibnamefont
  {Millot}}, \bibinfo {author} {\bibfnamefont {P.~M.}\ \bibnamefont
  {Celliers}}, \bibinfo {author} {\bibfnamefont {P.~A.}\ \bibnamefont
  {Sterne}}, \bibinfo {author} {\bibfnamefont {L.~X.}\ \bibnamefont
  {Benedict}}, \bibinfo {author} {\bibfnamefont {A.~A.}\ \bibnamefont
  {Correa}}, \bibinfo {author} {\bibfnamefont {S.}~\bibnamefont {Hamel}},
  \bibinfo {author} {\bibfnamefont {S.~J.}\ \bibnamefont {Ali}},  \emph
  {et~al.},\ }\href@noop {} {\bibfield  {journal} {\bibinfo  {journal} {Phys.
  Rev. B}\ }\textbf {\bibinfo {volume} {97}},\ \bibinfo {pages} {144108}
  (\bibinfo {year} {2018})}\BibitemShut {NoStop}%
\bibitem [{\citenamefont {Hopkins}\ \emph {et~al.}(2019)\citenamefont
  {Hopkins}, \citenamefont {LePape}, \citenamefont {Divol}, \citenamefont
  {Pak}, \citenamefont {Dewald} \emph {et~al.}}]{Hopkins2019}%
  \BibitemOpen
  \bibfield  {author} {\bibinfo {author} {\bibfnamefont {L.~B.}\ \bibnamefont
  {Hopkins}}, \bibinfo {author} {\bibfnamefont {S.}~\bibnamefont {LePape}},
  \bibinfo {author} {\bibfnamefont {L.}~\bibnamefont {Divol}}, \bibinfo
  {author} {\bibfnamefont {A.}~\bibnamefont {Pak}}, \bibinfo {author}
  {\bibfnamefont {E.}~\bibnamefont {Dewald}},  \emph {et~al.},\ }\href@noop {}
  {\bibfield  {journal} {\bibinfo  {journal} {Plasma Phys. Control. Fusion}\
  }\textbf {\bibinfo {volume} {61}},\ \bibinfo {pages} {014023} (\bibinfo
  {year} {2019})}\BibitemShut {NoStop}%
\bibitem [{\citenamefont {Kritcher}\ \emph {et~al.}(2021)\citenamefont
  {Kritcher}, \citenamefont {Zylstra}, \citenamefont {Callahan}, \citenamefont
  {Hurricane} \emph {et~al.}}]{Kritcher2021}%
  \BibitemOpen
  \bibfield  {author} {\bibinfo {author} {\bibfnamefont {A.~L.}\ \bibnamefont
  {Kritcher}}, \bibinfo {author} {\bibfnamefont {A.~B.}\ \bibnamefont
  {Zylstra}}, \bibinfo {author} {\bibfnamefont {D.~A.}\ \bibnamefont
  {Callahan}}, \bibinfo {author} {\bibfnamefont {O.~A.}\ \bibnamefont
  {Hurricane}},  \emph {et~al.},\ }\href@noop {} {\bibfield  {journal}
  {\bibinfo  {journal} {Phys. Plasmas}\ }\textbf {\bibinfo {volume} {28}},\
  \bibinfo {pages} {072706} (\bibinfo {year} {2021})}\BibitemShut {NoStop}%
\bibitem [{\citenamefont {Casey}\ \emph {et~al.}(2021)\citenamefont {Casey},
  \citenamefont {MacGowan}, \citenamefont {Sater}, \citenamefont {Zylstra},
  \citenamefont {Landen}, \citenamefont {Milovich}, \citenamefont {Hurricane},
  \citenamefont {Kritcher} \emph {et~al.}}]{Casey2021}%
  \BibitemOpen
  \bibfield  {author} {\bibinfo {author} {\bibfnamefont {D.~T.}\ \bibnamefont
  {Casey}}, \bibinfo {author} {\bibfnamefont {B.~J.}\ \bibnamefont {MacGowan}},
  \bibinfo {author} {\bibfnamefont {J.~D.}\ \bibnamefont {Sater}}, \bibinfo
  {author} {\bibfnamefont {A.~B.}\ \bibnamefont {Zylstra}}, \bibinfo {author}
  {\bibfnamefont {O.~L.}\ \bibnamefont {Landen}}, \bibinfo {author}
  {\bibfnamefont {J.}~\bibnamefont {Milovich}}, \bibinfo {author}
  {\bibfnamefont {O.~A.}\ \bibnamefont {Hurricane}}, \bibinfo {author}
  {\bibfnamefont {A.~L.}\ \bibnamefont {Kritcher}},  \emph {et~al.},\
  }\href@noop {} {\bibfield  {journal} {\bibinfo  {journal} {Phys. Rev. Lett.}\
  }\textbf {\bibinfo {volume} {126}},\ \bibinfo {pages} {25002} (\bibinfo
  {year} {2021})}\BibitemShut {NoStop}%
\bibitem [{\citenamefont {Tollefson}(2021)}]{Tollefson2021}%
  \BibitemOpen
  \bibfield  {author} {\bibinfo {author} {\bibfnamefont {J.}~\bibnamefont
  {Tollefson}},\ }\href@noop {} {\bibfield  {journal} {\bibinfo  {journal}
  {Nature}\ }\textbf {\bibinfo {volume} {597}},\ \bibinfo {pages} {163}
  (\bibinfo {year} {2021})}\BibitemShut {NoStop}%
\bibitem [{\citenamefont {Duffy}\ and\ \citenamefont
  {Smith}(2019)}]{Duffy2019}%
  \BibitemOpen
  \bibfield  {author} {\bibinfo {author} {\bibfnamefont {T.~S.}\ \bibnamefont
  {Duffy}}\ and\ \bibinfo {author} {\bibfnamefont {R.~F.}\ \bibnamefont
  {Smith}},\ }\href@noop {} {\bibfield  {journal} {\bibinfo  {journal} {Front.
  Earth Sci.}\ }\textbf {\bibinfo {volume} {7}},\ \bibinfo {pages} {1}
  (\bibinfo {year} {2019})}\BibitemShut {NoStop}%
\bibitem [{\citenamefont {Sinars}\ \emph {et~al.}(2020)\citenamefont {Sinars},
  \citenamefont {Sweeney}, \citenamefont {Alexander}, \citenamefont
  {Ampleford}, \citenamefont {Ao} \emph {et~al.}}]{Sinars2020}%
  \BibitemOpen
  \bibfield  {author} {\bibinfo {author} {\bibfnamefont {D.~B.}\ \bibnamefont
  {Sinars}}, \bibinfo {author} {\bibfnamefont {M.~A.}\ \bibnamefont {Sweeney}},
  \bibinfo {author} {\bibfnamefont {C.~S.}\ \bibnamefont {Alexander}}, \bibinfo
  {author} {\bibfnamefont {D.~J.}\ \bibnamefont {Ampleford}}, \bibinfo {author}
  {\bibfnamefont {T.}~\bibnamefont {Ao}},  \emph {et~al.},\ }\href@noop {}
  {\bibfield  {journal} {\bibinfo  {journal} {Phys. Plasmas}\ }\textbf
  {\bibinfo {volume} {27}},\ \bibinfo {pages} {070501} (\bibinfo {year}
  {2020})}\BibitemShut {NoStop}%
\bibitem [{\citenamefont {McMahon}(2020)}]{McMahon2020}%
  \BibitemOpen
  \bibfield  {author} {\bibinfo {author} {\bibfnamefont {M.~I.}\ \bibnamefont
  {McMahon}},\ }in\ \href@noop {} {\emph {\bibinfo {booktitle} {Synchrotron
  Light Sources Free. Lasers}}},\ \bibinfo {editor} {edited by\ \bibinfo
  {editor} {\bibfnamefont {E.~J.}\ \bibnamefont {Jaeschke}}, \bibinfo {editor}
  {\bibfnamefont {S.}~\bibnamefont {Khan}}, \bibinfo {editor} {\bibfnamefont
  {J.~R.}\ \bibnamefont {Schneider}}, \ and\ \bibinfo {editor} {\bibfnamefont
  {J.~B.}\ \bibnamefont {Hastings}}}\ (\bibinfo  {publisher} {Springer
  International Publishing},\ \bibinfo {year} {2020})\ pp.\ \bibinfo {pages}
  {1857--1896}\BibitemShut {NoStop}%
\bibitem [{\citenamefont {Seddon}\ \emph {et~al.}(2017)\citenamefont {Seddon},
  \citenamefont {Clarke}, \citenamefont {Dunning}, \citenamefont
  {Masciovecchio}, \citenamefont {Milne}, \citenamefont {Parmigiani},
  \citenamefont {Rugg}, \citenamefont {Spence}, \citenamefont {Thompson},
  \citenamefont {Ueda}, \citenamefont {Vinko}, \citenamefont {Wark},\ and\
  \citenamefont {Wurth}}]{Seddon2017}%
  \BibitemOpen
  \bibfield  {author} {\bibinfo {author} {\bibfnamefont {E.~A.}\ \bibnamefont
  {Seddon}}, \bibinfo {author} {\bibfnamefont {J.~A.}\ \bibnamefont {Clarke}},
  \bibinfo {author} {\bibfnamefont {D.~J.}\ \bibnamefont {Dunning}}, \bibinfo
  {author} {\bibfnamefont {C.}~\bibnamefont {Masciovecchio}}, \bibinfo {author}
  {\bibfnamefont {C.~J.}\ \bibnamefont {Milne}}, \bibinfo {author}
  {\bibfnamefont {F.}~\bibnamefont {Parmigiani}}, \bibinfo {author}
  {\bibfnamefont {D.}~\bibnamefont {Rugg}}, \bibinfo {author} {\bibfnamefont
  {J.~C.~H.}\ \bibnamefont {Spence}}, \bibinfo {author} {\bibfnamefont {N.~R.}\
  \bibnamefont {Thompson}}, \bibinfo {author} {\bibfnamefont {K.}~\bibnamefont
  {Ueda}}, \bibinfo {author} {\bibfnamefont {S.~M.}\ \bibnamefont {Vinko}},
  \bibinfo {author} {\bibfnamefont {J.~S.}\ \bibnamefont {Wark}}, \ and\
  \bibinfo {author} {\bibfnamefont {W.}~\bibnamefont {Wurth}},\ }\href@noop {}
  {\bibfield  {journal} {\bibinfo  {journal} {Reports Prog. Phys.}\ }\textbf
  {\bibinfo {volume} {80}},\ \bibinfo {pages} {115901} (\bibinfo {year}
  {2017})}\BibitemShut {NoStop}%
\bibitem [{\citenamefont {Pavlovskii}(1971)}]{Pavlovskii1971}%
  \BibitemOpen
  \bibfield  {author} {\bibinfo {author} {\bibfnamefont {M.~N.}\ \bibnamefont
  {Pavlovskii}},\ }\href@noop {} {\bibfield  {journal} {\bibinfo  {journal}
  {Sov. Phys. - Solid State}\ }\textbf {\bibinfo {volume} {13}},\ \bibinfo
  {pages} {741} (\bibinfo {year} {1971})}\BibitemShut {NoStop}%
\bibitem [{\citenamefont {Kondo}\ and\ \citenamefont
  {Ahrens}(1983)}]{Kondo1983}%
  \BibitemOpen
  \bibfield  {author} {\bibinfo {author} {\bibfnamefont {K.-I.}\ \bibnamefont
  {Kondo}}\ and\ \bibinfo {author} {\bibfnamefont {T.~J.}\ \bibnamefont
  {Ahrens}},\ }\href@noop {} {\bibfield  {journal} {\bibinfo  {journal}
  {Geophys. Res. Lett.}\ }\textbf {\bibinfo {volume} {10}},\ \bibinfo {pages}
  {281} (\bibinfo {year} {1983})}\BibitemShut {NoStop}%
\bibitem [{\citenamefont {McWilliams}\ \emph {et~al.}(2010)\citenamefont
  {McWilliams}, \citenamefont {Eggert}, \citenamefont {Hicks}, \citenamefont
  {Bradley}, \citenamefont {Celliers}, \citenamefont {Spaulding}, \citenamefont
  {Boehly}, \citenamefont {Collins},\ and\ \citenamefont
  {Jeanloz}}]{McWilliams2010}%
  \BibitemOpen
  \bibfield  {author} {\bibinfo {author} {\bibfnamefont {R.~S.}\ \bibnamefont
  {McWilliams}}, \bibinfo {author} {\bibfnamefont {J.~H.}\ \bibnamefont
  {Eggert}}, \bibinfo {author} {\bibfnamefont {D.~G.}\ \bibnamefont {Hicks}},
  \bibinfo {author} {\bibfnamefont {D.~K.}\ \bibnamefont {Bradley}}, \bibinfo
  {author} {\bibfnamefont {P.~M.}\ \bibnamefont {Celliers}}, \bibinfo {author}
  {\bibfnamefont {D.~K.}\ \bibnamefont {Spaulding}}, \bibinfo {author}
  {\bibfnamefont {T.~R.}\ \bibnamefont {Boehly}}, \bibinfo {author}
  {\bibfnamefont {G.~W.}\ \bibnamefont {Collins}}, \ and\ \bibinfo {author}
  {\bibfnamefont {R.}~\bibnamefont {Jeanloz}},\ }\href@noop {} {\bibfield
  {journal} {\bibinfo  {journal} {Phys. Rev. B}\ }\textbf {\bibinfo {volume}
  {81}},\ \bibinfo {pages} {27} (\bibinfo {year} {2010})}\BibitemShut {NoStop}%
\bibitem [{\citenamefont {Lang}\ \emph {et~al.}(2018)\citenamefont {Lang},
  \citenamefont {Winey},\ and\ \citenamefont {Gupta}}]{Lang2018}%
  \BibitemOpen
  \bibfield  {author} {\bibinfo {author} {\bibfnamefont {J.~M.}\ \bibnamefont
  {Lang}}, \bibinfo {author} {\bibfnamefont {J.~M.}\ \bibnamefont {Winey}}, \
  and\ \bibinfo {author} {\bibfnamefont {Y.~M.}\ \bibnamefont {Gupta}},\
  }\href@noop {} {\bibfield  {journal} {\bibinfo  {journal} {Phys. Rev. B}\
  }\textbf {\bibinfo {volume} {97}},\ \bibinfo {pages} {104106} (\bibinfo
  {year} {2018})}\BibitemShut {NoStop}%
\bibitem [{\citenamefont {Katagiri}\ \emph {et~al.}(2020)\citenamefont
  {Katagiri}, \citenamefont {Ozaki}, \citenamefont {Umeda}, \citenamefont
  {Irifune}, \citenamefont {Kamimura}, \citenamefont {Miyanishi}, \citenamefont
  {Sano}, \citenamefont {Sekine},\ and\ \citenamefont {Kodama}}]{Katagiri2020}%
  \BibitemOpen
  \bibfield  {author} {\bibinfo {author} {\bibfnamefont {K.}~\bibnamefont
  {Katagiri}}, \bibinfo {author} {\bibfnamefont {N.}~\bibnamefont {Ozaki}},
  \bibinfo {author} {\bibfnamefont {Y.}~\bibnamefont {Umeda}}, \bibinfo
  {author} {\bibfnamefont {T.}~\bibnamefont {Irifune}}, \bibinfo {author}
  {\bibfnamefont {N.}~\bibnamefont {Kamimura}}, \bibinfo {author}
  {\bibfnamefont {K.}~\bibnamefont {Miyanishi}}, \bibinfo {author}
  {\bibfnamefont {T.}~\bibnamefont {Sano}}, \bibinfo {author} {\bibfnamefont
  {T.}~\bibnamefont {Sekine}}, \ and\ \bibinfo {author} {\bibfnamefont
  {R.}~\bibnamefont {Kodama}},\ }\href@noop {} {\bibfield  {journal} {\bibinfo
  {journal} {Phys. Rev. Lett.}\ }\textbf {\bibinfo {volume} {125}},\ \bibinfo
  {pages} {185701} (\bibinfo {year} {2020})}\BibitemShut {NoStop}%
\bibitem [{\citenamefont {Winey}\ \emph {et~al.}(2020)\citenamefont {Winey},
  \citenamefont {Knudson},\ and\ \citenamefont {Gupta}}]{Winey2020}%
  \BibitemOpen
  \bibfield  {author} {\bibinfo {author} {\bibfnamefont {J.~M.}\ \bibnamefont
  {Winey}}, \bibinfo {author} {\bibfnamefont {M.~D.}\ \bibnamefont {Knudson}},
  \ and\ \bibinfo {author} {\bibfnamefont {Y.~M.}\ \bibnamefont {Gupta}},\
  }\href@noop {} {\bibfield  {journal} {\bibinfo  {journal} {Phys. Rev. B}\
  }\textbf {\bibinfo {volume} {101}},\ \bibinfo {pages} {184105} (\bibinfo
  {year} {2020})}\BibitemShut {NoStop}%
\bibitem [{\citenamefont {MacDonald}\ \emph {et~al.}(2020)\citenamefont
  {MacDonald}, \citenamefont {McBride}, \citenamefont {Galtier}, \citenamefont
  {Gauthier}, \citenamefont {Granados}, \citenamefont {Kraus}, \citenamefont
  {Krygier}, \citenamefont {Levitan}, \citenamefont {MacKinnon}, \citenamefont
  {Nam}, \citenamefont {Schumaker}, \citenamefont {Sun}, \citenamefont {van
  Driel}, \citenamefont {Vorberger}, \citenamefont {Xing}, \citenamefont
  {Drake}, \citenamefont {Glenzer},\ and\ \citenamefont
  {Fletcher}}]{MacDonald2020}%
  \BibitemOpen
  \bibfield  {author} {\bibinfo {author} {\bibfnamefont {M.~J.}\ \bibnamefont
  {MacDonald}}, \bibinfo {author} {\bibfnamefont {E.~E.}\ \bibnamefont
  {McBride}}, \bibinfo {author} {\bibfnamefont {E.}~\bibnamefont {Galtier}},
  \bibinfo {author} {\bibfnamefont {M.}~\bibnamefont {Gauthier}}, \bibinfo
  {author} {\bibfnamefont {E.}~\bibnamefont {Granados}}, \bibinfo {author}
  {\bibfnamefont {D.}~\bibnamefont {Kraus}}, \bibinfo {author} {\bibfnamefont
  {A.}~\bibnamefont {Krygier}}, \bibinfo {author} {\bibfnamefont {A.~L.}\
  \bibnamefont {Levitan}}, \bibinfo {author} {\bibfnamefont {A.~J.}\
  \bibnamefont {MacKinnon}}, \bibinfo {author} {\bibfnamefont {I.}~\bibnamefont
  {Nam}}, \bibinfo {author} {\bibfnamefont {W.}~\bibnamefont {Schumaker}},
  \bibinfo {author} {\bibfnamefont {P.}~\bibnamefont {Sun}}, \bibinfo {author}
  {\bibfnamefont {T.~B.}\ \bibnamefont {van Driel}}, \bibinfo {author}
  {\bibfnamefont {J.}~\bibnamefont {Vorberger}}, \bibinfo {author}
  {\bibfnamefont {Z.}~\bibnamefont {Xing}}, \bibinfo {author} {\bibfnamefont
  {R.~P.}\ \bibnamefont {Drake}}, \bibinfo {author} {\bibfnamefont {S.~H.}\
  \bibnamefont {Glenzer}}, \ and\ \bibinfo {author} {\bibfnamefont {L.~B.}\
  \bibnamefont {Fletcher}},\ }\href@noop {} {\bibfield  {journal} {\bibinfo
  {journal} {Appl. Phys. Lett.}\ }\textbf {\bibinfo {volume} {116}},\ \bibinfo
  {pages} {234104} (\bibinfo {year} {2020})}\BibitemShut {NoStop}%
\bibitem [{\citenamefont {Brygoo}\ \emph {et~al.}(2007)\citenamefont {Brygoo},
  \citenamefont {Henry}, \citenamefont {Loubeyre}, \citenamefont {Eggert},
  \citenamefont {Koenig}, \citenamefont {Loupias}, \citenamefont
  {Benuzzi-Mounaix},\ and\ \citenamefont {{Rabec Le Gloahec}}}]{Brygoo2007}%
  \BibitemOpen
  \bibfield  {author} {\bibinfo {author} {\bibfnamefont {S.}~\bibnamefont
  {Brygoo}}, \bibinfo {author} {\bibfnamefont {E.}~\bibnamefont {Henry}},
  \bibinfo {author} {\bibfnamefont {P.}~\bibnamefont {Loubeyre}}, \bibinfo
  {author} {\bibfnamefont {J.}~\bibnamefont {Eggert}}, \bibinfo {author}
  {\bibfnamefont {M.}~\bibnamefont {Koenig}}, \bibinfo {author} {\bibfnamefont
  {B.}~\bibnamefont {Loupias}}, \bibinfo {author} {\bibfnamefont
  {A.}~\bibnamefont {Benuzzi-Mounaix}}, \ and\ \bibinfo {author} {\bibfnamefont
  {M.}~\bibnamefont {{Rabec Le Gloahec}}},\ }\href@noop {} {\bibfield
  {journal} {\bibinfo  {journal} {Nat. Mater.}\ }\textbf {\bibinfo {volume}
  {6}},\ \bibinfo {pages} {274} (\bibinfo {year} {2007})}\BibitemShut {NoStop}%
\bibitem [{\citenamefont {Knudson}\ \emph {et~al.}(2008)\citenamefont
  {Knudson}, \citenamefont {Desjarlais},\ and\ \citenamefont
  {Dolan}}]{Knudson2008}%
  \BibitemOpen
  \bibfield  {author} {\bibinfo {author} {\bibfnamefont {M.~D.}\ \bibnamefont
  {Knudson}}, \bibinfo {author} {\bibfnamefont {M.~P.}\ \bibnamefont
  {Desjarlais}}, \ and\ \bibinfo {author} {\bibfnamefont {D.~H.}\ \bibnamefont
  {Dolan}},\ }\href@noop {} {\bibfield  {journal} {\bibinfo  {journal}
  {Science}\ }\textbf {\bibinfo {volume} {322}},\ \bibinfo {pages} {1822}
  (\bibinfo {year} {2008})}\BibitemShut {NoStop}%
\bibitem [{\citenamefont {Eggert}\ \emph {et~al.}(2010)\citenamefont {Eggert},
  \citenamefont {Hicks}, \citenamefont {Celliers}, \citenamefont {Bradley},
  \citenamefont {McWilliams}, \citenamefont {Jeanloz}, \citenamefont {Miller},
  \citenamefont {Boehly},\ and\ \citenamefont {Collins}}]{Eggert2009}%
  \BibitemOpen
  \bibfield  {author} {\bibinfo {author} {\bibfnamefont {J.~H.}\ \bibnamefont
  {Eggert}}, \bibinfo {author} {\bibfnamefont {D.~G.}\ \bibnamefont {Hicks}},
  \bibinfo {author} {\bibfnamefont {P.~M.}\ \bibnamefont {Celliers}}, \bibinfo
  {author} {\bibfnamefont {D.~K.}\ \bibnamefont {Bradley}}, \bibinfo {author}
  {\bibfnamefont {R.~S.}\ \bibnamefont {McWilliams}}, \bibinfo {author}
  {\bibfnamefont {R.}~\bibnamefont {Jeanloz}}, \bibinfo {author} {\bibfnamefont
  {J.~E.}\ \bibnamefont {Miller}}, \bibinfo {author} {\bibfnamefont {T.~R.}\
  \bibnamefont {Boehly}}, \ and\ \bibinfo {author} {\bibfnamefont {G.~W.}\
  \bibnamefont {Collins}},\ }\href@noop {} {\bibfield  {journal} {\bibinfo
  {journal} {Nat. Phys.}\ }\textbf {\bibinfo {volume} {6}},\ \bibinfo {pages}
  {40} (\bibinfo {year} {2010})}\BibitemShut {NoStop}%
\bibitem [{\citenamefont {Gregor}\ \emph {et~al.}(2017)\citenamefont {Gregor},
  \citenamefont {Fratanduono}, \citenamefont {McCoy}, \citenamefont {Polsin},
  \citenamefont {Sorce}, \citenamefont {Rygg}, \citenamefont {Collins},
  \citenamefont {Braun}, \citenamefont {Celliers}, \citenamefont {Eggert},
  \citenamefont {Meyerhofer},\ and\ \citenamefont {Boehly}}]{Gregor2017}%
  \BibitemOpen
  \bibfield  {author} {\bibinfo {author} {\bibfnamefont {M.~C.}\ \bibnamefont
  {Gregor}}, \bibinfo {author} {\bibfnamefont {D.~E.}\ \bibnamefont
  {Fratanduono}}, \bibinfo {author} {\bibfnamefont {C.~A.}\ \bibnamefont
  {McCoy}}, \bibinfo {author} {\bibfnamefont {D.~N.}\ \bibnamefont {Polsin}},
  \bibinfo {author} {\bibfnamefont {A.}~\bibnamefont {Sorce}}, \bibinfo
  {author} {\bibfnamefont {J.~R.}\ \bibnamefont {Rygg}}, \bibinfo {author}
  {\bibfnamefont {G.~W.}\ \bibnamefont {Collins}}, \bibinfo {author}
  {\bibfnamefont {T.}~\bibnamefont {Braun}}, \bibinfo {author} {\bibfnamefont
  {P.~M.}\ \bibnamefont {Celliers}}, \bibinfo {author} {\bibfnamefont {J.~H.}\
  \bibnamefont {Eggert}}, \bibinfo {author} {\bibfnamefont {D.~D.}\
  \bibnamefont {Meyerhofer}}, \ and\ \bibinfo {author} {\bibfnamefont {T.~R.}\
  \bibnamefont {Boehly}},\ }\href@noop {} {\bibfield  {journal} {\bibinfo
  {journal} {Phys. Rev. B}\ }\textbf {\bibinfo {volume} {95}},\ \bibinfo
  {pages} {144114} (\bibinfo {year} {2017})}\BibitemShut {NoStop}%
\bibitem [{\citenamefont {Millot}\ \emph {et~al.}(2020)\citenamefont {Millot},
  \citenamefont {Sterne}, \citenamefont {Eggert}, \citenamefont {Hamel},
  \citenamefont {Marshall},\ and\ \citenamefont {Celliers}}]{Millot2020}%
  \BibitemOpen
  \bibfield  {author} {\bibinfo {author} {\bibfnamefont {M.}~\bibnamefont
  {Millot}}, \bibinfo {author} {\bibfnamefont {P.~A.}\ \bibnamefont {Sterne}},
  \bibinfo {author} {\bibfnamefont {J.~H.}\ \bibnamefont {Eggert}}, \bibinfo
  {author} {\bibfnamefont {S.}~\bibnamefont {Hamel}}, \bibinfo {author}
  {\bibfnamefont {M.~C.}\ \bibnamefont {Marshall}}, \ and\ \bibinfo {author}
  {\bibfnamefont {P.~M.}\ \bibnamefont {Celliers}},\ }\href@noop {} {\bibfield
  {journal} {\bibinfo  {journal} {Phys. Plasmas}\ }\textbf {\bibinfo {volume}
  {27}},\ \bibinfo {pages} {102711} (\bibinfo {year} {2020})}\BibitemShut
  {NoStop}%
\bibitem [{\citenamefont {Lazicki}\ \emph {et~al.}(2021)\citenamefont
  {Lazicki}, \citenamefont {McGonegle}, \citenamefont {Rygg}, \citenamefont
  {Braun}, \citenamefont {Swift} \emph {et~al.}}]{Lazicki2021}%
  \BibitemOpen
  \bibfield  {author} {\bibinfo {author} {\bibfnamefont {A.}~\bibnamefont
  {Lazicki}}, \bibinfo {author} {\bibfnamefont {D.}~\bibnamefont {McGonegle}},
  \bibinfo {author} {\bibfnamefont {J.~R.}\ \bibnamefont {Rygg}}, \bibinfo
  {author} {\bibfnamefont {D.~G.}\ \bibnamefont {Braun}}, \bibinfo {author}
  {\bibfnamefont {D.~C.}\ \bibnamefont {Swift}},  \emph {et~al.},\ }\href@noop
  {} {\bibfield  {journal} {\bibinfo  {journal} {Nature}\ }\textbf {\bibinfo
  {volume} {589}},\ \bibinfo {pages} {532} (\bibinfo {year}
  {2021})}\BibitemShut {NoStop}%
\bibitem [{\citenamefont {Scandolo}\ \emph {et~al.}(1996)\citenamefont
  {Scandolo}, \citenamefont {Chiarotti},\ and\ \citenamefont
  {Tosatti}}]{Scandolo1996}%
  \BibitemOpen
  \bibfield  {author} {\bibinfo {author} {\bibfnamefont {S.}~\bibnamefont
  {Scandolo}}, \bibinfo {author} {\bibfnamefont {G.}~\bibnamefont {Chiarotti}},
  \ and\ \bibinfo {author} {\bibfnamefont {E.}~\bibnamefont {Tosatti}},\
  }\href@noop {} {\bibfield  {journal} {\bibinfo  {journal} {Phys. Rev. B}\
  }\textbf {\bibinfo {volume} {53}},\ \bibinfo {pages} {5051} (\bibinfo {year}
  {1996})}\BibitemShut {NoStop}%
\bibitem [{\citenamefont {Grumbach}\ and\ \citenamefont
  {Martin}(1996)}]{Grumbach1996}%
  \BibitemOpen
  \bibfield  {author} {\bibinfo {author} {\bibfnamefont {M.~P.}\ \bibnamefont
  {Grumbach}}\ and\ \bibinfo {author} {\bibfnamefont {R.~M.}\ \bibnamefont
  {Martin}},\ }\href@noop {} {\bibfield  {journal} {\bibinfo  {journal} {Phys.
  Rev. B}\ }\textbf {\bibinfo {volume} {54}},\ \bibinfo {pages} {15730}
  (\bibinfo {year} {1996})}\BibitemShut {NoStop}%
\bibitem [{\citenamefont {Wang}\ \emph {et~al.}(2005)\citenamefont {Wang},
  \citenamefont {Scandolo},\ and\ \citenamefont {Car}}]{Wang2005}%
  \BibitemOpen
  \bibfield  {author} {\bibinfo {author} {\bibfnamefont {X.}~\bibnamefont
  {Wang}}, \bibinfo {author} {\bibfnamefont {S.}~\bibnamefont {Scandolo}}, \
  and\ \bibinfo {author} {\bibfnamefont {R.}~\bibnamefont {Car}},\ }\href@noop
  {} {\bibfield  {journal} {\bibinfo  {journal} {Phys. Rev. Lett.}\ }\textbf
  {\bibinfo {volume} {95}},\ \bibinfo {pages} {1} (\bibinfo {year}
  {2005})}\BibitemShut {NoStop}%
\bibitem [{\citenamefont {Correa}\ \emph {et~al.}(2006)\citenamefont {Correa},
  \citenamefont {Bonev},\ and\ \citenamefont {Galli}}]{Correa2006}%
  \BibitemOpen
  \bibfield  {author} {\bibinfo {author} {\bibfnamefont {A.~A.}\ \bibnamefont
  {Correa}}, \bibinfo {author} {\bibfnamefont {S.~A.}\ \bibnamefont {Bonev}}, \
  and\ \bibinfo {author} {\bibfnamefont {G.}~\bibnamefont {Galli}},\
  }\href@noop {} {\bibfield  {journal} {\bibinfo  {journal} {Proc. Natl. Acad.
  Sci. U. S. A.}\ }\textbf {\bibinfo {volume} {103}},\ \bibinfo {pages} {1204}
  (\bibinfo {year} {2006})}\BibitemShut {NoStop}%
\bibitem [{\citenamefont {Correa}\ \emph {et~al.}(2008)\citenamefont {Correa},
  \citenamefont {Benedict}, \citenamefont {Young}, \citenamefont {Schwegler},\
  and\ \citenamefont {Bonev}}]{Correa2008}%
  \BibitemOpen
  \bibfield  {author} {\bibinfo {author} {\bibfnamefont {A.}~\bibnamefont
  {Correa}}, \bibinfo {author} {\bibfnamefont {L.}~\bibnamefont {Benedict}},
  \bibinfo {author} {\bibfnamefont {D.}~\bibnamefont {Young}}, \bibinfo
  {author} {\bibfnamefont {E.}~\bibnamefont {Schwegler}}, \ and\ \bibinfo
  {author} {\bibfnamefont {S.~A.}\ \bibnamefont {Bonev}},\ }\href@noop {}
  {\bibfield  {journal} {\bibinfo  {journal} {Phys. Rev. B}\ }\textbf {\bibinfo
  {volume} {78}},\ \bibinfo {pages} {25} (\bibinfo {year} {2008})}\BibitemShut
  {NoStop}%
\bibitem [{\citenamefont {Sun}\ \emph {et~al.}(2009)\citenamefont {Sun},
  \citenamefont {Klug},\ and\ \citenamefont {Marto\^{n}\'{a}k}}]{Sun2009}%
  \BibitemOpen
  \bibfield  {author} {\bibinfo {author} {\bibfnamefont {J.}~\bibnamefont
  {Sun}}, \bibinfo {author} {\bibfnamefont {D.~D.}\ \bibnamefont {Klug}}, \
  and\ \bibinfo {author} {\bibfnamefont {R.}~\bibnamefont {Marto\^{n}\'{a}k}},\
  }\href@noop {} {\bibfield  {journal} {\bibinfo  {journal} {J. Chem. Phys.}\
  }\textbf {\bibinfo {volume} {130}},\ \bibinfo {pages} {194512} (\bibinfo
  {year} {2009})}\BibitemShut {NoStop}%
\bibitem [{\citenamefont {Martinez-Canales}\ \emph {et~al.}(2012)\citenamefont
  {Martinez-Canales}, \citenamefont {Pickard},\ and\ \citenamefont
  {Needs}}]{Martinez-Canales2012}%
  \BibitemOpen
  \bibfield  {author} {\bibinfo {author} {\bibfnamefont {M.}~\bibnamefont
  {Martinez-Canales}}, \bibinfo {author} {\bibfnamefont {C.~J.}\ \bibnamefont
  {Pickard}}, \ and\ \bibinfo {author} {\bibfnamefont {R.~J.}\ \bibnamefont
  {Needs}},\ }\href@noop {} {\bibfield  {journal} {\bibinfo  {journal} {Phys.
  Rev. Lett.}\ }\textbf {\bibinfo {volume} {108}},\ \bibinfo {pages} {45704}
  (\bibinfo {year} {2012})}\BibitemShut {NoStop}%
\bibitem [{\citenamefont {Benedict}\ \emph {et~al.}(2014)\citenamefont
  {Benedict}, \citenamefont {Driver}, \citenamefont {Hamel}, \citenamefont
  {Militzer}, \citenamefont {Qi}, \citenamefont {Correa}, \citenamefont
  {Saul},\ and\ \citenamefont {Schwegler}}]{Benedict2014}%
  \BibitemOpen
  \bibfield  {author} {\bibinfo {author} {\bibfnamefont {L.~X.}\ \bibnamefont
  {Benedict}}, \bibinfo {author} {\bibfnamefont {K.~P.}\ \bibnamefont
  {Driver}}, \bibinfo {author} {\bibfnamefont {S.}~\bibnamefont {Hamel}},
  \bibinfo {author} {\bibfnamefont {B.}~\bibnamefont {Militzer}}, \bibinfo
  {author} {\bibfnamefont {T.}~\bibnamefont {Qi}}, \bibinfo {author}
  {\bibfnamefont {A.~A.}\ \bibnamefont {Correa}}, \bibinfo {author}
  {\bibfnamefont {A.}~\bibnamefont {Saul}}, \ and\ \bibinfo {author}
  {\bibfnamefont {E.}~\bibnamefont {Schwegler}},\ }\href@noop {} {\bibfield
  {journal} {\bibinfo  {journal} {Phys. Rev. B}\ }\textbf {\bibinfo {volume}
  {89}},\ \bibinfo {pages} {224109} (\bibinfo {year} {2014})}\BibitemShut
  {NoStop}%
\bibitem [{\citenamefont {Thompson}\ \emph {et~al.}(2022)\citenamefont
  {Thompson}, \citenamefont {Aktulga}, \citenamefont {Berger}, \citenamefont
  {Bolintineanu}, \citenamefont {Brown}, \citenamefont {Crozier}, \citenamefont
  {{in 't Veld}}, \citenamefont {Kohlmeyer}, \citenamefont {Moore},
  \citenamefont {Nguyen}, \citenamefont {Shan}, \citenamefont {Stevens},
  \citenamefont {Tranchida}, \citenamefont {Trott},\ and\ \citenamefont
  {Plimpton}}]{Thompson2022}%
  \BibitemOpen
  \bibfield  {author} {\bibinfo {author} {\bibfnamefont {A.~P.}\ \bibnamefont
  {Thompson}}, \bibinfo {author} {\bibfnamefont {H.~M.}\ \bibnamefont
  {Aktulga}}, \bibinfo {author} {\bibfnamefont {R.}~\bibnamefont {Berger}},
  \bibinfo {author} {\bibfnamefont {D.~S.}\ \bibnamefont {Bolintineanu}},
  \bibinfo {author} {\bibfnamefont {W.~M.}\ \bibnamefont {Brown}}, \bibinfo
  {author} {\bibfnamefont {P.~S.}\ \bibnamefont {Crozier}}, \bibinfo {author}
  {\bibfnamefont {P.~J.}\ \bibnamefont {{in 't Veld}}}, \bibinfo {author}
  {\bibfnamefont {A.}~\bibnamefont {Kohlmeyer}}, \bibinfo {author}
  {\bibfnamefont {S.~G.}\ \bibnamefont {Moore}}, \bibinfo {author}
  {\bibfnamefont {T.~D.}\ \bibnamefont {Nguyen}}, \bibinfo {author}
  {\bibfnamefont {R.}~\bibnamefont {Shan}}, \bibinfo {author} {\bibfnamefont
  {M.~J.}\ \bibnamefont {Stevens}}, \bibinfo {author} {\bibfnamefont
  {J.}~\bibnamefont {Tranchida}}, \bibinfo {author} {\bibfnamefont
  {C.}~\bibnamefont {Trott}}, \ and\ \bibinfo {author} {\bibfnamefont {S.~J.}\
  \bibnamefont {Plimpton}},\ }\href@noop {} {\bibfield  {journal} {\bibinfo
  {journal} {Comput. Phys. Commun.}\ }\textbf {\bibinfo {volume} {271}},\
  \bibinfo {pages} {108171} (\bibinfo {year} {2022})}\BibitemShut {NoStop}%
\bibitem [{\citenamefont {Brenner}\ \emph {et~al.}(2002)\citenamefont
  {Brenner}, \citenamefont {Shenderova}, \citenamefont {Harrison},
  \citenamefont {Stuart}, \citenamefont {Ni},\ and\ \citenamefont
  {Sinnott}}]{Brenner2002}%
  \BibitemOpen
  \bibfield  {author} {\bibinfo {author} {\bibfnamefont {D.~W.}\ \bibnamefont
  {Brenner}}, \bibinfo {author} {\bibfnamefont {O.~A.}\ \bibnamefont
  {Shenderova}}, \bibinfo {author} {\bibfnamefont {J.~A.}\ \bibnamefont
  {Harrison}}, \bibinfo {author} {\bibfnamefont {S.~J.}\ \bibnamefont
  {Stuart}}, \bibinfo {author} {\bibfnamefont {B.}~\bibnamefont {Ni}}, \ and\
  \bibinfo {author} {\bibfnamefont {S.~B.}\ \bibnamefont {Sinnott}},\
  }\href@noop {} {\bibfield  {journal} {\bibinfo  {journal} {J. Phys. Condens.
  Matter}\ }\textbf {\bibinfo {volume} {14}},\ \bibinfo {pages} {783} (\bibinfo
  {year} {2002})}\BibitemShut {NoStop}%
\bibitem [{\citenamefont {Stuart}\ \emph {et~al.}(2000)\citenamefont {Stuart},
  \citenamefont {Tutein},\ and\ \citenamefont {Harrison}}]{Stuart2000}%
  \BibitemOpen
  \bibfield  {author} {\bibinfo {author} {\bibfnamefont {S.~J.}\ \bibnamefont
  {Stuart}}, \bibinfo {author} {\bibfnamefont {A.~B.}\ \bibnamefont {Tutein}},
  \ and\ \bibinfo {author} {\bibfnamefont {J.~A.}\ \bibnamefont {Harrison}},\
  }\href {\doibase 10.1063/1.481208} {\bibfield  {journal} {\bibinfo  {journal}
  {J. Chem. Phys}\ }\textbf {\bibinfo {volume} {112}},\ \bibinfo {pages} {6472}
  (\bibinfo {year} {2000})}\BibitemShut {NoStop}%
\bibitem [{\citenamefont {Los}\ and\ \citenamefont {Fasolino}(2003)}]{Los2003}%
  \BibitemOpen
  \bibfield  {author} {\bibinfo {author} {\bibfnamefont {J.~H.}\ \bibnamefont
  {Los}}\ and\ \bibinfo {author} {\bibfnamefont {A.}~\bibnamefont {Fasolino}},\
  }\href@noop {} {\bibfield  {journal} {\bibinfo  {journal} {Phys. Rev. B}\
  }\textbf {\bibinfo {volume} {68}},\ \bibinfo {pages} {024107} (\bibinfo
  {year} {2003})}\BibitemShut {NoStop}%
\bibitem [{\citenamefont {Pastewka}\ \emph {et~al.}(2008)\citenamefont
  {Pastewka}, \citenamefont {Pou}, \citenamefont {P{\'{e}}rez}, \citenamefont
  {Gumbsch},\ and\ \citenamefont {Moseler}}]{Pastewka2008}%
  \BibitemOpen
  \bibfield  {author} {\bibinfo {author} {\bibfnamefont {L.}~\bibnamefont
  {Pastewka}}, \bibinfo {author} {\bibfnamefont {P.}~\bibnamefont {Pou}},
  \bibinfo {author} {\bibfnamefont {R.}~\bibnamefont {P{\'{e}}rez}}, \bibinfo
  {author} {\bibfnamefont {P.}~\bibnamefont {Gumbsch}}, \ and\ \bibinfo
  {author} {\bibfnamefont {M.}~\bibnamefont {Moseler}},\ }\href@noop {}
  {\bibfield  {journal} {\bibinfo  {journal} {Phys. Rev. B}\ }\textbf {\bibinfo
  {volume} {78}},\ \bibinfo {pages} {161402} (\bibinfo {year}
  {2008})}\BibitemShut {NoStop}%
\bibitem [{\citenamefont {Srinivasan}\ \emph {et~al.}(2015)\citenamefont
  {Srinivasan}, \citenamefont {van Duin},\ and\ \citenamefont
  {Ganesh}}]{Srinivasan2015}%
  \BibitemOpen
  \bibfield  {author} {\bibinfo {author} {\bibfnamefont {S.~G.}\ \bibnamefont
  {Srinivasan}}, \bibinfo {author} {\bibfnamefont {A.~C.~T.}\ \bibnamefont {van
  Duin}}, \ and\ \bibinfo {author} {\bibfnamefont {P.}~\bibnamefont {Ganesh}},\
  }\href@noop {} {\bibfield  {journal} {\bibinfo  {journal} {J. Phys. Chem. A}\
  }\textbf {\bibinfo {volume} {119}},\ \bibinfo {pages} {571} (\bibinfo {year}
  {2015})}\BibitemShut {NoStop}%
\bibitem [{\citenamefont {Oleynik}\ \emph {et~al.}(2008)\citenamefont
  {Oleynik}, \citenamefont {Landerville}, \citenamefont {Zybin}, \citenamefont
  {Elert},\ and\ \citenamefont {White}}]{Oleynik2008}%
  \BibitemOpen
  \bibfield  {author} {\bibinfo {author} {\bibfnamefont {I.~I.}\ \bibnamefont
  {Oleynik}}, \bibinfo {author} {\bibfnamefont {A.~C.}\ \bibnamefont
  {Landerville}}, \bibinfo {author} {\bibfnamefont {S.~V.}\ \bibnamefont
  {Zybin}}, \bibinfo {author} {\bibfnamefont {M.~L.}\ \bibnamefont {Elert}}, \
  and\ \bibinfo {author} {\bibfnamefont {C.~T.}\ \bibnamefont {White}},\
  }\href@noop {} {\bibfield  {journal} {\bibinfo  {journal} {Phys. Rev. B}\
  }\textbf {\bibinfo {volume} {78}},\ \bibinfo {pages} {180101} (\bibinfo
  {year} {2008})}\BibitemShut {NoStop}%
\bibitem [{\citenamefont {Perriot}\ \emph {et~al.}(2013)\citenamefont
  {Perriot}, \citenamefont {Gu}, \citenamefont {Lin}, \citenamefont
  {Zhakhovsky},\ and\ \citenamefont {Oleynik}}]{Perriot2013}%
  \BibitemOpen
  \bibfield  {author} {\bibinfo {author} {\bibfnamefont {R.}~\bibnamefont
  {Perriot}}, \bibinfo {author} {\bibfnamefont {X.}~\bibnamefont {Gu}},
  \bibinfo {author} {\bibfnamefont {Y.}~\bibnamefont {Lin}}, \bibinfo {author}
  {\bibfnamefont {V.~V.}\ \bibnamefont {Zhakhovsky}}, \ and\ \bibinfo {author}
  {\bibfnamefont {I.~I.}\ \bibnamefont {Oleynik}},\ }\href@noop {} {\bibfield
  {journal} {\bibinfo  {journal} {Phys. Rev. B}\ }\textbf {\bibinfo {volume}
  {88}},\ \bibinfo {pages} {064101} (\bibinfo {year} {2013})}\BibitemShut
  {NoStop}%
\bibitem [{\citenamefont {Behler}\ and\ \citenamefont
  {Parrinello}(2007)}]{Behler2007}%
  \BibitemOpen
  \bibfield  {author} {\bibinfo {author} {\bibfnamefont {J.}~\bibnamefont
  {Behler}}\ and\ \bibinfo {author} {\bibfnamefont {M.}~\bibnamefont
  {Parrinello}},\ }\href@noop {} {\bibfield  {journal} {\bibinfo  {journal}
  {Phys. Rev. Lett.}\ }\textbf {\bibinfo {volume} {98}},\ \bibinfo {pages} {1}
  (\bibinfo {year} {2007})}\BibitemShut {NoStop}%
\bibitem [{\citenamefont {Bart{\'{o}}k}\ \emph {et~al.}(2010)\citenamefont
  {Bart{\'{o}}k}, \citenamefont {Payne}, \citenamefont {Kondor},\ and\
  \citenamefont {Cs{\'{a}}nyi}}]{Bartok2010}%
  \BibitemOpen
  \bibfield  {author} {\bibinfo {author} {\bibfnamefont {A.~P.}\ \bibnamefont
  {Bart{\'{o}}k}}, \bibinfo {author} {\bibfnamefont {M.~C.}\ \bibnamefont
  {Payne}}, \bibinfo {author} {\bibfnamefont {R.}~\bibnamefont {Kondor}}, \
  and\ \bibinfo {author} {\bibfnamefont {G.}~\bibnamefont {Cs{\'{a}}nyi}},\
  }\href@noop {} {\bibfield  {journal} {\bibinfo  {journal} {Phys. Rev. Lett.}\
  }\textbf {\bibinfo {volume} {104}},\ \bibinfo {pages} {136403} (\bibinfo
  {year} {2010})}\BibitemShut {NoStop}%
\bibitem [{\citenamefont {Friederich}\ \emph {et~al.}(2021)\citenamefont
  {Friederich}, \citenamefont {H{\"{a}}se}, \citenamefont {Proppe},\ and\
  \citenamefont {Aspuru-Guzik}}]{Friederich2021}%
  \BibitemOpen
  \bibfield  {author} {\bibinfo {author} {\bibfnamefont {P.}~\bibnamefont
  {Friederich}}, \bibinfo {author} {\bibfnamefont {F.}~\bibnamefont
  {H{\"{a}}se}}, \bibinfo {author} {\bibfnamefont {J.}~\bibnamefont {Proppe}},
  \ and\ \bibinfo {author} {\bibfnamefont {A.}~\bibnamefont {Aspuru-Guzik}},\
  }\href@noop {} {\bibfield  {journal} {\bibinfo  {journal} {Nat. Mater.}\
  }\textbf {\bibinfo {volume} {20}},\ \bibinfo {pages} {750} (\bibinfo {year}
  {2021})}\BibitemShut {NoStop}%
\bibitem [{\citenamefont {Thompson}\ \emph {et~al.}(2015)\citenamefont
  {Thompson}, \citenamefont {Swiler}, \citenamefont {Trott}, \citenamefont
  {Foiles},\ and\ \citenamefont {Tucker}}]{Thompson2015}%
  \BibitemOpen
  \bibfield  {author} {\bibinfo {author} {\bibfnamefont {A.~P.}\ \bibnamefont
  {Thompson}}, \bibinfo {author} {\bibfnamefont {L.~P.}\ \bibnamefont
  {Swiler}}, \bibinfo {author} {\bibfnamefont {C.~R.}\ \bibnamefont {Trott}},
  \bibinfo {author} {\bibfnamefont {S.~M.}\ \bibnamefont {Foiles}}, \ and\
  \bibinfo {author} {\bibfnamefont {G.~J.}\ \bibnamefont {Tucker}},\
  }\href@noop {} {\bibfield  {journal} {\bibinfo  {journal} {J. Comput. Phys.}\
  }\textbf {\bibinfo {volume} {285}},\ \bibinfo {pages} {316} (\bibinfo {year}
  {2015})}\BibitemShut {NoStop}%
\bibitem [{\citenamefont {Wood}\ and\ \citenamefont
  {Thompson}(2018)}]{Wood2018}%
  \BibitemOpen
  \bibfield  {author} {\bibinfo {author} {\bibfnamefont {M.~A.}\ \bibnamefont
  {Wood}}\ and\ \bibinfo {author} {\bibfnamefont {A.~P.}\ \bibnamefont
  {Thompson}},\ }\href@noop {} {\bibfield  {journal} {\bibinfo  {journal} {J.
  Chem. Phys.}\ }\textbf {\bibinfo {volume} {148}},\ \bibinfo {pages} {241721}
  (\bibinfo {year} {2018})}\BibitemShut {NoStop}%
\bibitem [{\citenamefont {Shapeev}(2015)}]{Shapeev2015}%
  \BibitemOpen
  \bibfield  {author} {\bibinfo {author} {\bibfnamefont {A.~V.}\ \bibnamefont
  {Shapeev}},\ }\href@noop {} {\bibfield  {journal} {\bibinfo  {journal}
  {Multiscale Model. Simul.}\ }\textbf {\bibinfo {volume} {14}},\ \bibinfo
  {pages} {1} (\bibinfo {year} {2015})}\BibitemShut {NoStop}%
\bibitem [{\citenamefont {Drautz}(2019)}]{Drautz2019}%
  \BibitemOpen
  \bibfield  {author} {\bibinfo {author} {\bibfnamefont {R.}~\bibnamefont
  {Drautz}},\ }\href@noop {} {\bibfield  {journal} {\bibinfo  {journal} {Phys.
  Rev. B}\ }\textbf {\bibinfo {volume} {99}},\ \bibinfo {pages} {1} (\bibinfo
  {year} {2019})}\BibitemShut {NoStop}%
\bibitem [{\citenamefont {Zhang}\ \emph {et~al.}(2018)\citenamefont {Zhang},
  \citenamefont {Han}, \citenamefont {Wang}, \citenamefont {Car},\ and\
  \citenamefont {Weinan}}]{Zhang2018}%
  \BibitemOpen
  \bibfield  {author} {\bibinfo {author} {\bibfnamefont {L.}~\bibnamefont
  {Zhang}}, \bibinfo {author} {\bibfnamefont {J.}~\bibnamefont {Han}}, \bibinfo
  {author} {\bibfnamefont {H.}~\bibnamefont {Wang}}, \bibinfo {author}
  {\bibfnamefont {R.}~\bibnamefont {Car}}, \ and\ \bibinfo {author}
  {\bibfnamefont {E.}~\bibnamefont {Weinan}},\ }\href@noop {} {\bibfield
  {journal} {\bibinfo  {journal} {Phys. Rev. Lett.}\ }\textbf {\bibinfo
  {volume} {120}},\ \bibinfo {pages} {143001} (\bibinfo {year}
  {2018})}\BibitemShut {NoStop}%
\bibitem [{\citenamefont {Khaliullin}\ \emph {et~al.}(2010)\citenamefont
  {Khaliullin}, \citenamefont {Eshet}, \citenamefont {K{\"{u}}hne},
  \citenamefont {Behler},\ and\ \citenamefont {Parrinello}}]{Khaliullin2010}%
  \BibitemOpen
  \bibfield  {author} {\bibinfo {author} {\bibfnamefont {R.~Z.}\ \bibnamefont
  {Khaliullin}}, \bibinfo {author} {\bibfnamefont {H.}~\bibnamefont {Eshet}},
  \bibinfo {author} {\bibfnamefont {T.~D.}\ \bibnamefont {K{\"{u}}hne}},
  \bibinfo {author} {\bibfnamefont {J.}~\bibnamefont {Behler}}, \ and\ \bibinfo
  {author} {\bibfnamefont {M.}~\bibnamefont {Parrinello}},\ }\href@noop {}
  {\bibfield  {journal} {\bibinfo  {journal} {Phys. Rev. B}\ }\textbf {\bibinfo
  {volume} {81}},\ \bibinfo {pages} {18} (\bibinfo {year} {2010})}\BibitemShut
  {NoStop}%
\bibitem [{\citenamefont {Rowe}\ \emph {et~al.}(2020)\citenamefont {Rowe},
  \citenamefont {Deringer}, \citenamefont {Gasparotto}, \citenamefont
  {Cs{\'{a}}nyi},\ and\ \citenamefont {Michaelides}}]{Rowe2020}%
  \BibitemOpen
  \bibfield  {author} {\bibinfo {author} {\bibfnamefont {P.}~\bibnamefont
  {Rowe}}, \bibinfo {author} {\bibfnamefont {V.~L.}\ \bibnamefont {Deringer}},
  \bibinfo {author} {\bibfnamefont {P.}~\bibnamefont {Gasparotto}}, \bibinfo
  {author} {\bibfnamefont {G.}~\bibnamefont {Cs{\'{a}}nyi}}, \ and\ \bibinfo
  {author} {\bibfnamefont {A.}~\bibnamefont {Michaelides}},\ }\href@noop {}
  {\bibfield  {journal} {\bibinfo  {journal} {J. Chem. Phys.}\ }\textbf
  {\bibinfo {volume} {153}},\ \bibinfo {pages} {034702} (\bibinfo {year}
  {2020})}\BibitemShut {NoStop}%
\bibitem [{\citenamefont {Zuo}\ \emph {et~al.}(2020)\citenamefont {Zuo},
  \citenamefont {Chen}, \citenamefont {Li}, \citenamefont {Deng}, \citenamefont
  {Chen}, \citenamefont {Behler}, \citenamefont {Cs{\'a}nyi}, \citenamefont
  {Shapeev}, \citenamefont {Thompson}, \citenamefont {Wood},\ and\
  \citenamefont {Ong}}]{Ong2020}%
  \BibitemOpen
  \bibfield  {author} {\bibinfo {author} {\bibfnamefont {Y.}~\bibnamefont
  {Zuo}}, \bibinfo {author} {\bibfnamefont {C.}~\bibnamefont {Chen}}, \bibinfo
  {author} {\bibfnamefont {X.}~\bibnamefont {Li}}, \bibinfo {author}
  {\bibfnamefont {Z.}~\bibnamefont {Deng}}, \bibinfo {author} {\bibfnamefont
  {Y.}~\bibnamefont {Chen}}, \bibinfo {author} {\bibfnamefont {J.}~\bibnamefont
  {Behler}}, \bibinfo {author} {\bibfnamefont {G.}~\bibnamefont {Cs{\'a}nyi}},
  \bibinfo {author} {\bibfnamefont {A.~V.}\ \bibnamefont {Shapeev}}, \bibinfo
  {author} {\bibfnamefont {A.~P.}\ \bibnamefont {Thompson}}, \bibinfo {author}
  {\bibfnamefont {M.~A.}\ \bibnamefont {Wood}}, \ and\ \bibinfo {author}
  {\bibfnamefont {S.~P.}\ \bibnamefont {Ong}},\ }\href@noop {} {\bibfield
  {journal} {\bibinfo  {journal} {J. Phys. Chem. A}\ }\textbf {\bibinfo
  {volume} {124}},\ \bibinfo {pages} {731} (\bibinfo {year}
  {2020})}\BibitemShut {NoStop}%
\bibitem [{S_m()}]{S_material}%
  \BibitemOpen
  \href@noop {} {}\bibinfo {note} {See Supplemental Material at
  http://link.aps.org/ supplemental/ for details on (1) SNAP training database;
  (2) SNAP machine learning development workflow; (3) Accuracy assessment of
  SNAP training; (4) SNAP validation: Radial Distribution Functions of solid
  and liquid phases along diamond and BC8 melting lines.}\BibitemShut {Stop}%
\bibitem [{\citenamefont {Belonoshko}(1994)}]{Belonoshko1994}%
  \BibitemOpen
  \bibfield  {author} {\bibinfo {author} {\bibfnamefont {A.~B.}\ \bibnamefont
  {Belonoshko}},\ }\href {\doibase 10.1016/0016-7037(94)90265-8} {\bibfield
  {journal} {\bibinfo  {journal} {Geochimica et Cosmochimica Acta}\ }\textbf
  {\bibinfo {volume} {58}},\ \bibinfo {pages} {4039} (\bibinfo {year}
  {1994})}\BibitemShut {NoStop}%
\bibitem [{\citenamefont {Williams}\ \emph {et~al.}(2022)\citenamefont
  {Williams}, \citenamefont {Nguyen-Cong}, \citenamefont {Willman},\ and\
  \citenamefont {Oleynik}}]{Williams2022}%
  \BibitemOpen
  \bibfield  {author} {\bibinfo {author} {\bibfnamefont {A.~S.}\ \bibnamefont
  {Williams}}, \bibinfo {author} {\bibfnamefont {K.}~\bibnamefont
  {Nguyen-Cong}}, \bibinfo {author} {\bibfnamefont {J.~T.}\ \bibnamefont
  {Willman}}, \ and\ \bibinfo {author} {\bibfnamefont {I.~I.}\ \bibnamefont
  {Oleynik}},\ }\href@noop {} {\bibfield  {journal} {\bibinfo  {journal} {Phys.
  Chem. Lett.}\ ,\ \bibinfo {pages} {submitted for publication}} (\bibinfo
  {year} {2022})}\BibitemShut {NoStop}%
\bibitem [{fit()}]{fitsnap}%
  \BibitemOpen
  \href@noop {} {}\bibinfo {note} {FitSNAP: Software for generating SNAP
  machine learning interatomic potentials:
  http://github.com/FitSNAP}\BibitemShut {NoStop}%
\bibitem [{\citenamefont {Adams}(2021)}]{Adams}%
  \BibitemOpen
  \bibfield  {author} {\bibinfo {author} {\bibfnamefont {B.}~\bibnamefont
  {Adams}},\ }\href@noop {} {\emph {\bibinfo {title} {Dakota, A Multilevel
  Parallel Object-Oriented Framework for Design Optimization, Parameter
  Estimation, Uncertainty Quantification, and Sensitivity Analysis: Version
  6.14 User's Manual}}} (\bibinfo {year} {2021})\BibitemShut {NoStop}%
\bibitem [{\citenamefont {Nguyen-Cong}\ \emph {et~al.}(2021)\citenamefont
  {Nguyen-Cong}, \citenamefont {Willman}, \citenamefont {Moore}, \citenamefont
  {Belonoshko}, \citenamefont {Gayatri}, \citenamefont {Weinberg},
  \citenamefont {Wood}, \citenamefont {Thompson},\ and\ \citenamefont
  {Oleynik}}]{Nguyen-Cong2021}%
  \BibitemOpen
  \bibfield  {author} {\bibinfo {author} {\bibfnamefont {K.}~\bibnamefont
  {Nguyen-Cong}}, \bibinfo {author} {\bibfnamefont {J.~T.}\ \bibnamefont
  {Willman}}, \bibinfo {author} {\bibfnamefont {S.~G.}\ \bibnamefont {Moore}},
  \bibinfo {author} {\bibfnamefont {A.~B.}\ \bibnamefont {Belonoshko}},
  \bibinfo {author} {\bibfnamefont {R.}~\bibnamefont {Gayatri}}, \bibinfo
  {author} {\bibfnamefont {E.}~\bibnamefont {Weinberg}}, \bibinfo {author}
  {\bibfnamefont {M.~A.}\ \bibnamefont {Wood}}, \bibinfo {author}
  {\bibfnamefont {A.~P.}\ \bibnamefont {Thompson}}, \ and\ \bibinfo {author}
  {\bibfnamefont {I.~I.}\ \bibnamefont {Oleynik}},\ }\href@noop {} {\bibfield
  {journal} {\bibinfo  {journal} {Proc. Int. Conf. High Perform. Comput.
  Networking, Storage Anal.}\ ,\ \bibinfo {pages} {1}} (\bibinfo {year}
  {2021})}\BibitemShut {NoStop}%
\end{thebibliography}%

\end{document}